\documentclass[10pt, conference]{IEEEtran}
\pdfoutput=1

\newif\ifcameraready

\camerareadytrue 

\usepackage[utf8]{inputenc}
\usepackage[hidelinks]{hyperref}
\usepackage{amsmath}
\usepackage{booktabs}
\usepackage{cite}
\usepackage{cleveref}
\usepackage{listings}
\usepackage{makecell}
\usepackage{multirow}
\usepackage{pgfplots}
\usepackage{pifont}
\usepackage{siunitx}
\usepackage{tabularx}
\usepackage{tcolorbox}
\usepackage{todonotes}
\tikzset{notestyleraw/.append style={align=justify}}
\usepackage{url}
\usepackage{verbatim}
\usepackage{xspace}
\usepackage{array}
\newcolumntype{P}[1]{>{\centering\arraybackslash}p{#1}}
\usepackage{color, colortbl}
\usepackage{textcomp}
\usepackage{tcolorbox}
\tcbuselibrary{breakable}

\usetikzlibrary{arrows, shadows}

\crefname{lstlisting}{listing}{listings}
\Crefname{lstlisting}{Listing}{Listings}

\newcommand{\rosa}{\textsc{Rosa}\xspace}
\newcommand{\rosarum}{\textsc{Rosarum}\xspace}
\newcommand\BackdoorsTotal{17\xspace}
\newcommand\BackdoorsAuthentic{7\xspace}
\newcommand\BackdoorsSynthetic{10\xspace}
\newcommand\ReplicationPackageRepoURL{\url{https://zenodo.org/records/14724251}\xspace}
\newcommand\RosaToolAnonRepoURL{\url{https://github.com/icse25-938/rosa}\xspace}
\newcommand\BackdoorDatasetAnonRepoURL{\url{https://github.com/icse25-938/rosarum}\xspace}
\newcommand{\myparagraph}[1]{\smallskip \noindent \textbf{#1.}} 

\newcommand\TotalRuns{\num{180}\xspace}
\newcommand\SuccessfulRuns{\num{156}\xspace}
\newcommand\SuccessfulRunsPercentage{\num{87}\%\xspace}
\newcommand\FailedRuns{\num{24}\xspace}

\newcommand\BackdoorsWithAllSuccessfulRuns{\num{11}\xspace}
\newcommand\BackdoorsWithAllSuccessfulRunsPercentage{\num{65}\%\xspace}
\newcommand\RosaDetectedBackdoors{17\xspace}
\newcommand\BackdoorsWithHalfSuccessfulRuns{\num{17}\xspace}
\newcommand\BackdoorsWithHalfSuccessfulRunsPercentage{\num{100}\%\xspace}
\newcommand\RosaAverageDetectionTime{1h30m35s\xspace}
\newcommand\RosaAverageDetectionTimeClean{1h30\xspace}
\newcommand\BackdoorsWithAverageDetectionTimeBelowOneHour{\num{9}\xspace}
\newcommand\BackdoorsWithAverageDetectionTimeBelowOneHourPercentage{\num{53}\%\xspace}
\newcommand\BackdoorsWithAverageDetectionTimeBelowFiveHours{\num{16}\xspace}
\newcommand\BackdoorsWithAverageDetectionTimeBelowFiveHoursPercentage{\num{94}\%\xspace}
\newcommand\RosaAverageInputs{\num{7}\xspace}
\newcommand\BackdoorsWithAverageOneInput{\num{4}\xspace}
\newcommand\BackdoorsWithAverageOneInputPercentage{\num{24}\%\xspace}
\newcommand\BackdoorsWithAverageTenInputs{\num{13}\xspace}
\newcommand\BackdoorsWithAverageTenInputsPercentage{\num{76}\%\xspace}
\newcommand\BackdoorsWithAverageMoreThanTenInputs{\num{4}\xspace}
\newcommand\BackdoorsWithAverageMoreThanTenInputsPercentage{\num{24}\%\xspace}
\newcommand\StringerDetectedBackdoors{4\xspace}
\newcommand\StringerDetectedBackdoorsPercentage{\num{24}\%\xspace}
\newcommand\StringerAverageTimesFasterThanRosa{1,928\xspace}
\newcommand\StringerAverageInputs{\num{308}\xspace}
\newcommand\StringerAverageTimesMoreInputsThanRosa{44\xspace}

\newcommand\BackdoorContaminationRunsPercentage{\num{5.56}\%\xspace}

\newcommand\BestPhaseOneDuration{1 minute\xspace}
 
\title{
    \vspace{-1.6cm}
    \begin{flushright}
        \href{https://www.acm.org/publications/policies/artifact-review-and-badging-current}{
            \includegraphics[height=40pt]{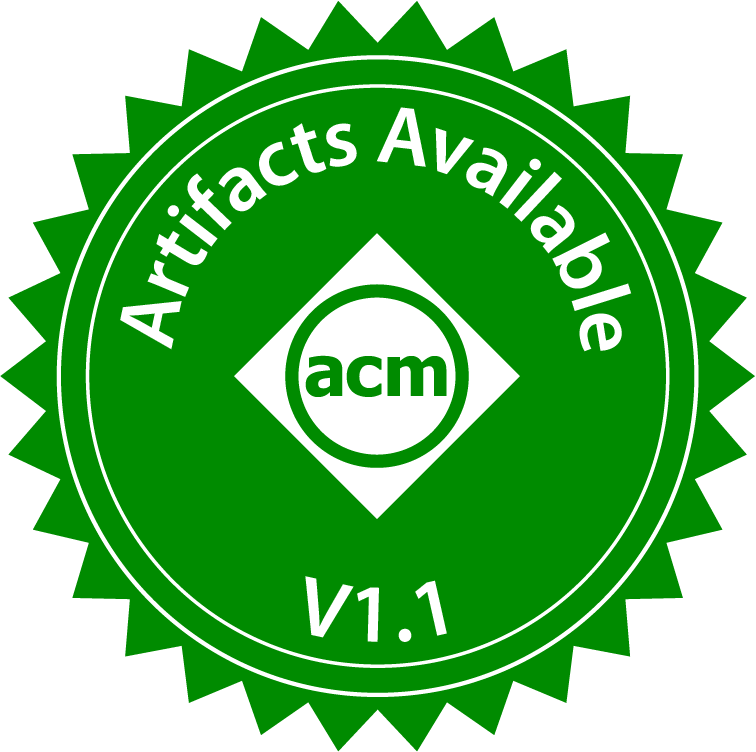}
            \includegraphics[height=40pt]{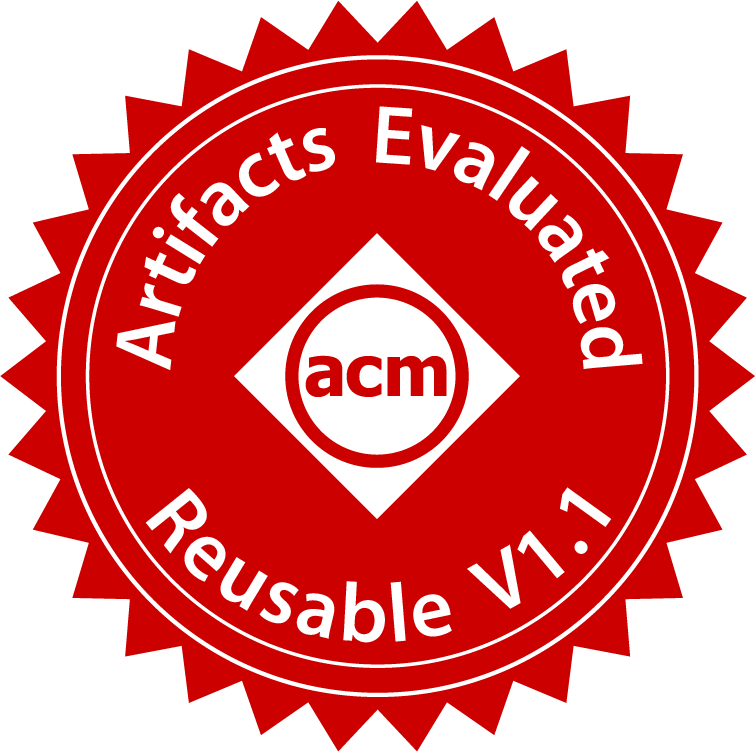}
        }
    \end{flushright}

    \rosa: Finding Backdoors with Fuzzing
}

\ifcameraready  
\author{
   \IEEEauthorblockN{Dimitri Kokkonis, Michaël Marcozzi, Emilien Decoux}
   \IEEEauthorblockA{
    Université Paris-Saclay, CEA, List\\
    Paris-Saclay, France\\
    first.last@cea.fr
   }
   \and
   \IEEEauthorblockN{Stefano Zacchiroli}
   \IEEEauthorblockA{
       LTCI, Télécom Paris, Institut Polytechnique de Paris\\
       Palaiseau, France\\
       stefano.zacchiroli@telecom-paris.fr
   }
}
\else 
\author{
    \IEEEauthorblockN{Anonymous Author(s)}
    \IEEEauthorblockA{
        Some affiliation\\
        Some place\\
        anony@mo.us
    }
} 
\fi

\begin{document}

\maketitle

\begin{abstract}

A code-level backdoor is a hidden access, programmed and concealed within the code of a program.
    For instance, hard-coded credentials planted in the code of a file server application would enable  maliciously logging into all deployed instances of this application. Confirmed software supply-chain attacks have led to the injection of backdoors into popular open-source projects, and backdoors have been discovered in various router firmware. Manual code auditing for backdoors is  challenging and existing semi-automated approaches can handle only a limited scope of programs and backdoors, while requiring manual reverse-engineering of the audited (binary) program.
Graybox fuzzing (automated semi-randomized testing) has grown in popularity due to its success in discovering vulnerabilities and hence stands as a strong candidate for improved backdoor detection. However, current fuzzing knowledge does not offer any means to detect the triggering of a backdoor at runtime.

In this work we introduce \rosa, a novel approach (and tool) which combines a state-of-the-art fuzzer (AFL++) with a new metamorphic test oracle, capable of detecting runtime backdoor triggers. 
To facilitate the evaluation of \rosa, we have created \rosarum, the first openly available
    benchmark for assessing the detection of various backdoors in diverse programs. Experimental
    evaluation shows that \rosa has a level of robustness, speed and automation similar to
    classical fuzzing. It finds all 17 authentic or synthetic backdooors from \rosarum in 1h30 on
    average. Compared to existing detection tools, it can handle a diversity of backdoors and
    programs and it does not rely on manual reverse-engineering of the fuzzed binary code.  

\end{abstract}


\section{Introduction}
\label{sec:intro}

\myparagraph{Context} A \emph{code-level backdoor}~\cite{thomas-2018} is a hidden access,
programmed and concealed within the code of a program. It enables program users aware of the
backdoor to feed the program with a specific input value and thus trigger a privilege escalation within the program or gain undue access to underlying system resources. For example, hard-coded credentials planted in the code base of a file server application can enable  maliciously logging into all deployed instances of this application in the world. Confirmed software supply-chain attacks have led to the injection of backdoors into popular open-source projects, like PHP~\cite{ganz-2023}, ProFTPD~\cite{schuster-2013}, vsFTPd~\cite{evans-2011} and xz~\cite{CVE-2024-3094}. Backdoors have also been discovered in the binary firmware of popular network routers~\cite{toterhi-2018, lee-2013, craig-2013, benoist-vanderbeken-2015}. 

\emph{Graybox fuzzing}~\cite{godefroid-2020} is a form of automated program testing. It relies on a search-based approach~\cite{harman-2009} to generate test inputs automatically and on simple test oracles~\cite{barr-2015} (such as crash detection and code sanitizers~\cite{song-2019}) to detect failures automatically at runtime. Among advanced capabilities, modern fuzzing tools (or \emph{fuzzers}), like the community-maintained AFL++~\cite{fioraldi-2020}, are now equipped for testing binary-only programs~\cite{qemu-afl} and for efficiently exploring complex branching conditions in the tested code~\cite{laf-intel-2016, aschermann-2019}. These tools are currently attracting a lot of popularity and research efforts, notably because of their reported ability~\cite{fioraldi-2020} to discover software vulnerabilities in programs. 

\myparagraph{Problem} Addressing the threat of backdoors requires proper auditing of software dependencies and (binary) firmware.
Yet, this
necessitates a long and painstaking manual inspection of large amounts of code, so that auditing is
often not performed at all~\cite{thomas-stringer-2017}. While there has been some progress in
automating backdoor detection~\cite{schuster-2013, shoshitaishvili-2015, thomas-stringer-2017,
thomas-humidify-2017}, the state of the art still suffers from the limited scope of programs and
backdoors that can be handled. In addition, current detection tools still rely on manual
reverse-engineering of the vetted (binary) code.

\myparagraph{Goal and challenges} In this work, we aim at taking advantage of the capabilities of modern graybox fuzzers to automate backdoor detection. Fuzzing has indeed the potential to enable backdoor detection for a wide variety of programs and backdoors, with no manual code reverse-engineering.
Yet, while the current body of knowledge in fuzzing enables generating test inputs for a wide range of programs, it does not offer any means to detect the triggering of a backdoor at runtime. In addition, benchmarking backdoor detection capabilities over multiple programs and backdoors is difficult, as backdoor reports in the literature are scarce and often point to lost samples or undocumented binary firmware, running on obsolete and difficult-to-obtain appliances.

\myparagraph{Proposal} We introduce \rosa, a novel approach (and tool) which combines a state-of-the-art fuzzer (AFL++) with a new metamorphic test oracle~\cite{segura-2016}, capable of detecting backdoor triggers at runtime. The key intuition behind \rosa is that, for example, fuzzing a backdoored file server application with incorrect credentials should always cause \emph{similar observable reactions}; however, among the generated wrong credentials, the ones that trigger the backdoor will cause a \emph{different reaction}, enabling the \rosa oracle to detect them.

To facilitate the experimental evaluation of \rosa and its comparison with existing tools, we have created and made available the novel \rosarum benchmark, consisting of \BackdoorsAuthentic authentic backdoors, coupled with \BackdoorsSynthetic diverse synthetic backdoors inserted into a standard fuzzing benchmark~\cite{hazimeh-2020}. 

\myparagraph{Evaluation} We run 10 fuzzing campaigns, lasting 8 hours each, using \rosa on each
backdoor in the \rosarum benchmark. \rosa can detect all \RosaDetectedBackdoors backdoors in \RosaAverageDetectionTimeClean on average, demonstrating a level of
robustness and speed similar to vanilla AFL++ for classical bugs. The automation level is also
similar to AFL++, but \rosa may produce false positives that must then be semi-automatically
discarded. Yet, the required manual effort is limited to vetting (an average of \RosaAverageInputs) suspicious runtime behaviors detected while fuzzing, like the launching of a root shell. 
This level of performance primarily qualifies \rosa as a good candidate to increase automation during large-scale code auditing events, like before deploying router firmware or software dependencies in critical infrastructures.

We compare in
depth against \textsc{Stringer}, the only competing backdoor detection tool that is available and
working. As it relies on a simple static analysis, \textsc{Stringer} is faster than \rosa, but can
only detect \StringerDetectedBackdoors out the \BackdoorsTotal \rosarum backdoors and produces
\StringerAverageTimesMoreInputsThanRosa times more false positives.

\myparagraph{Contributions} Our main contributions are:
\begin{enumerate}
\item A new metamorphic oracle (based on a novel metamorphic relation and a fresh heuristic approach to find pairs of related inputs) which makes it possible, for the first time, to use graybox fuzzing as a means to detect code-level backdoors.
\item \rosa, an efficient backdoor detection method and tool, which complements a state-of-the-art fuzzer (AFL++) with our new metamorphic oracle. \rosa significantly improves the state of the art in backdoor detection, by (1) enabling the efficient discovery of a wider range of backdoors in a wider range of programs, compared to what existing approaches can do, and by (2) removing the need to manually reverse-engineer the analyzed (binary) code, which existing approaches still do require.
\item \rosarum, the first openly available benchmark for evaluating backdoor detection tools, as well as largest backdoor dataset ever used as per the state of the art.
\end{enumerate}

\ifcameraready  
    \myparagraph{Data availability statement} \rosa and \rosarum are available on GitHub and are
    archived on Software Heritage~\cite{rosa-repo-archive, rosarum-repo-archive}.
    A result replication package is available at \ReplicationPackageRepoURL.
\fi

\section{Background}
\label{sec:background}

\subsection{Code-level backdoors}
\label{subsec:backdoors}

\subsubsection{Definition and scope}

A backdoor in a fortified castle is an unprotected but concealed access. It enables those who are informed of its existence to circumvent all castle fortifications and enter without effort. By analogy, a \emph{backdoor in a computer system} is a hidden mechanism built by developers, able to grant undue privileges to users who are informed about it. Various instances of backdoors have been reported in diverse computer systems, like hardware backdoors in processors~\cite{lokhande-2019}, mathematical backdoors in cryptographic algorithms~\cite{kostyuk-2022} and data poisoning backdoors in machine learning models~\cite{chen-2017}. In this work, we focus on \emph{code-level backdoors}~\cite{thomas-2018} in classical software programs, like file server applications or command-line tools. In a nutshell, code-level backdoors are hidden features programmed and concealed within the program code base. They enable informed program users to leverage a specific \emph{key input value}, to trigger either a privilege escalation within the program or to get undue access to resources in the underlying system. For example, the key of a code-level backdoor can be a set of hard-coded credentials in the authentication module of a web server program~\cite{lee-2013} or a non-standard FTP command giving root access to a file server application's underlying operating system~\cite{schuster-2013}.

\subsubsection{Occurrence in the real world}

At least two types of software attacks in the real world have involved the secret injection of a code-level backdoor in a program. The first type is \emph{software supply-chain attacks}~\cite{ohm-2020,ladisa-2023}. The vast majority of modern software relies on thousands of third-party open-source or proprietary dependencies~\cite{kikas-2017, decan-2019-dep-networks}, and some of them come with backdoors. Injections of code-level backdoors in popular open-source software, such as PHP~\cite{ganz-2023}, ProFTPD~\cite{schuster-2013}, vsFTPd~\cite{evans-2011} and xz~\cite{CVE-2024-3094}, have been reported. In addition, dubious software developers may sell proprietary components infected by deliberate or unintentional backdoors (like a forgotten debug access). The second identified type of attacks involving backdoors relates to \emph{dubious Internet of Things (IoT) manufacturers}. IoT devices, like network or surveillance appliances, typically come with embedded proprietary software, called \emph{firmware}, driving their operations. Injecting a code-level backdoor in such firmware can enable compromising millions of devices worldwide, in possibly critical infrastructures. Firmware infections with backdoors have been reported for various types of routers~\cite{toterhi-2018,lee-2013,craig-2013,benoist-vanderbeken-2015}.

\subsubsection{Detection} 

Addressing the threat of code-level backdoors requires a proper auditing of software dependencies and IoT firmware, to ensure that they do not contain any such backdoor. Yet, in practice, such auditing is difficult, so that it is either not done or it requires a long and painstaking code inspection by a human expert~\cite{thomas-stringer-2017}. In particular, in the case of proprietary dependencies and IoT firmware, the code often comes in binary-only form, with no access to the source code and to the development logs, which are useful information to vet such software at a large scale. While there has been progress in automating code-level backdoor detection~\cite{schuster-2013, shoshitaishvili-2015, thomas-stringer-2017, thomas-humidify-2017}, the state of the art is still limited in terms of scope and level of automation, with no significant advancements in the field for more than seven years.

\subsection{Graybox and metamorphic fuzzing}
\label{subsec:fuzzing}

\subsubsection{Graybox fuzzing}

\emph{Fuzzing}~\cite{godefroid-2020} was originally
introduced~\cite{miller-1990} as a specific form of automated software
testing. It aimed at finding crashes in programs like UNIX utilities, by
feeding them with randomly generated test inputs. Nowadays, fuzzing is commonly
understood as a more general synonym of automated program testing. During a
fuzzing campaign, the program under test (PUT) is fed with a suite of test
inputs produced by an \emph{automated input generator}. In parallel, the
runtime behavior of the PUT with these inputs is analyzed for traces of
possible failures by an \emph{automated test oracle}. The uncovered failures
are symptoms of defects in the PUT, to be fixed before they cause harm or get
exploited in case they pose a security threat. Fuzzing tools (or
\emph{fuzzers}) mainly differ by the way their input generators and test oracles
work. In recent years, there has been a surge of interest in a specific family
of fuzzers, called \emph{graybox fuzzers}, which have been shown to be
successful for security vulnerability
detection~\cite{fioraldi-2020}.
\textbf{Graybox input generators} are typically based on the principles of search-based testing~\cite{harman-2009}, where the PUT's input space is explored using search heuristics, designed to maximize the part of the PUT code covered by the selected inputs. Compared to pure random input generators (blackbox fuzzing) and to those based on precise code analyses (whitebox fuzzing, a.k.a. symbolic execution~\cite{cadar-2008}), graybox generators aim at finding a sweet spot between (1) the ease and speed at which inputs can be generated, and (2) the ability to generate inputs achieving a high coverage of the PUT codebase. 
\textbf{Graybox oracles} usually rely on a lightweight mechanism, like crash detection, possibly coupled with code assertions in the PUT and sanitizers~\cite{song-2019}.

\subsubsection{The AFL++ graybox fuzzer}

American Fuzzy Lop (AFL) and its community-maintained successor \emph{AFL++}~\cite{zalewski-2016,fioraldi-2020} are some of the most used and forked graybox fuzzers. They rely on a \emph{mutation loop} that generates new inputs by randomly mutating (i.e., slightly modifying) some of the inputs generated during the previous iterations. More precisely, as the newly generated inputs are fed to the PUT, those that improve the coverage of neglected parts of the PUT code base are saved as \emph{seed inputs} (or \emph{seeds}). Only those seeds are then considered for mutation in the next iterations, possibly exploring even more the neglected parts of the PUT. The loop is bootstrapped with user-provided initial seeds, on which the first mutations are performed. In order to gather the code coverage data needed to guide this process, AFL++ injects additional code into the PUT. This \emph{instrumentation code} tracks and reports which parts of the PUT's \emph{original code} are covered during a run. Concretely, coverage is measured at the level of edges in the control-flow graph (CFG) of the PUT. Each edge is represented by a corresponding byte in a dedicated \textit{coverage map}. The instrumentation code zeroes the map at the start of the PUT execution and, each time the execution passes through a given edge, it increments the corresponding byte in the map.

\subsubsection{Binary fuzzing with AFL++}

AFL++ usually performs instrumentation while compiling the source code of the PUT, via a special
compilation pass plugged into a mainstream compiler. This pass injects the needed instructions in each PUT basic block, enabling the aforementioned edge coverage tracking. Yet, in this work, we aim at using AFL++ for detecting backdoors, which often requires vetting binary-only programs. For such situations, AFL++ now provides a slower binary fuzzing mode, where the PUT is run in an emulator---such as QEMU~\cite{bellard-2005}---since injecting the instrumentation directly into the binary is hard to perform robustly and accurately. The emulator tracks the coverage data on the fly, by approximating edges as jumps between the memory addresses in which it has loaded the PUT code.

\subsubsection{Guessing magic bytes with AFL++}
\label{subsubsec:magic-byte-fuzzing}
A historical limitation of graybox fuzzers is their difficulty to generate inputs able to traverse specific types of branching conditions, called \textit{magic byte comparisons}. An example of such a condition is
\begin{lstlisting}[language=C,basicstyle=\footnotesize\ttfamily,numbers=none,keywordstyle=\color{blue}\bfseries,]
if (input[3] == 0xdeadbeef) { /* Some code here */ }
\end{lstlisting}
where a part of the input is compared to the magic byte value \lstinline[language=c,basicstyle=\small\ttfamily]{0xdeadbeef}.
An old enough version of AFL++ would struggle to find inputs getting past this
\lstinline[language=c,basicstyle=\small\ttfamily,keywordstyle=\color{blue}\bfseries,]{if}
condition, as it would have to come up with the magic byte value by a series of random mutations
from the initial seed values. This becomes highly unlikely to achieve in reasonable time for long
enough magic byte sequences, making the fuzzer unable to test the whole part of the PUT code inside
the \lstinline[language=c,basicstyle=\small\ttfamily,keywordstyle=\color{blue}\bfseries,]{if}
condition. This is an important problem in the context of this work, as entire classes of
backdoors---such as hard-coded credentials---may rely on a magic-byte comparison---e.g., with
the hard-coded username or password---as a trigger. However, recent techniques that involve
splitting multi-byte comparisons into single-byte ones~\cite{laf-intel-2016}, or matching the
target bytes using lightweight taint tracking~\cite{aschermann-2019}, enable fuzzers to traverse magic byte comparisons much more efficiently. These techniques have been bundled into more recent versions of AFL++, making them credible candidates to detect backdoors. 

\subsubsection{Metamorphic oracles}

AFL++ relies on crash detection as its main oracle mechanism. As well-coded backdoors should not cause a crash when triggered by their key input value, they cannot be detected with such an oracle. A more sensitive oracle should thus be devised for AFL++ to perceive their triggering, and developing such an oracle is a core contribution of this work. Several families of sophisticated oracles~\cite{barr-2015} have already been proposed to detect complex forms of bugs and vulnerabilities. One such family is \emph{metamorphic oracles}~\cite{segura-2016}, based on \emph{metamorphic relations} that are expected to hold between pairs of inputs to a PUT. The principle of the oracle is then that any pair of generated inputs found violating the metamorphic relation is a trace of a PUT failure, to be further investigated. 
The oracle developed in this work to detect the triggering of backdoors is a form of a metamorphic oracle. Known successful uses of metamorphic oracles in fuzzing have notably enabled detecting intricate logic bugs in various complex, mature and large programs, like compilers, SQL database management systems and SMT solvers~\cite{yao-2021, emi-project, rigger-2020, lascu-2021, sqlancer}.

\section{Motivating example}
\label{sec:motivating-example}

\subsection{A ``hard-coded credentials'' backdoor in sudo}
The \textit{sudo} Unix command-line tool \cite{sudo} enables executing a given
command as a different (usually more privileged) user. For example, \lstinline[language=bash,basicstyle=\footnotesize\ttfamily,numbers=none,keywordstyle=\color{blue}\bfseries,morekeywords={sudo}]{echo PASSWORD | sudo -S -u alice CMD},
when run by an entitled user \lstinline[language=bash,basicstyle=\footnotesize\ttfamily,numbers=none,keywordstyle=\color{blue}\bfseries,morekeywords={sudo}]{bob}, allows them to run command \lstinline[language=bash,basicstyle=\footnotesize\ttfamily,numbers=none,keywordstyle=\color{blue}\bfseries,morekeywords={sudo}]{CMD} as user \lstinline[language=bash,basicstyle=\footnotesize\ttfamily,numbers=none,keywordstyle=\color{blue}\bfseries,morekeywords={sudo}]{alice}, provided that \lstinline[language=bash,basicstyle=\footnotesize\ttfamily,numbers=none,keywordstyle=\color{blue}\bfseries,morekeywords={sudo}]{PASSWORD} is the correct password for user \lstinline[language=bash,basicstyle=\footnotesize\ttfamily,numbers=none,keywordstyle=\color{blue}\bfseries,morekeywords={sudo}]{bob}.
If the password is indeed correct, sudo issues system calls to create a child process owned by \lstinline[language=bash,basicstyle=\footnotesize\ttfamily,numbers=none,keywordstyle=\color{blue}\bfseries,morekeywords={sudo}]{alice}, in which it executes \lstinline[language=bash,basicstyle=\footnotesize\ttfamily,numbers=none,keywordstyle=\color{blue}\bfseries,morekeywords={sudo}]{CMD}. Otherwise, sudo issues system calls to print an error message on the screen.

Let us now imagine that an attacker has injected the code-level backdoor from
\Cref{lst:sudo_backdoor} in sudo. This backdoor relies on a hard-coded credentials trigger (lines
5--8) which overwrites the result of the password and entitlement check from line 4. Regardless of which user executes sudo, impersonation will always succeed if they enter the password \lstinline[language=C,basicstyle=\footnotesize\ttfamily,numbers=none,keywordstyle=\color{blue}\bfseries,morekeywords={sudo}]{"let_me_in"}. This gives the attacker (and anyone informed of the key) full root access in any system containing the backdoored sudo.

\begin{lstlisting}[
    float,
    caption={
        Example of a ``hard-coded credentials'' backdoor in sudo.
    },
    captionpos=b,
    label=lst:sudo_backdoor,
    language=C,
    numbers=left,
    numberstyle=\scriptsize,
    basicstyle=\footnotesize\ttfamily,
    xleftmargin=1.65em,
    xrightmargin=0.35em,
    frame=single,
    keywordstyle=\color{blue}\bfseries,
    commentstyle=\color{red}
]
int verify_user(const struct sudoers_context* ctx 
  ,const char* password)
{
  int ret = ctx->verify(password);
  // --- Beginning of backdoor ---
  if (strcmp(password, "let_me_in") == 0) 
    {  ret = AUTH_SUCCESS;  }  
  // --- End of backdoor ---
  return ret;
}
\end{lstlisting}

\subsection{Detecting the backdoor with \rosa}
\label{subsec:detecting-with-rosa}

In order to understand how \rosa detects backdoors with a graybox fuzzer, we need to introduce the notion of \emph{input families} of a PUT.
Intuitively, the input values of a PUT can be classified into different families, where each family is a set of input values considered as similar in the PUT's specification, so that they result in close-by execution paths being taken in the PUT and similar effects on the PUT's environment.
\rosa then introduces a \emph{new metamorphic oracle}, relying on a new metamorphic relation, whose violations will be considered signs of possible backdoor presence: if two input values belong to the same {input family}, running the PUT on either of them should produce a similar effect on the PUT environment. 

Let us illustrate how input families enable detecting backdoors in the sudo backdoor example.
First, consider that \lstinline[breaklines=true,language=bash,basicstyle=\footnotesize\ttfamily,numbers=none,keywordstyle=\color{blue}\bfseries,morekeywords={sudo}]{echo PASSWORD | sudo -S -u alice CMD} is run by the user \lstinline[language=bash,basicstyle=\footnotesize\ttfamily,numbers=none,keywordstyle=\color{blue}\bfseries,morekeywords={sudo}]{bob} with a fixed \lstinline[language=bash,basicstyle=\footnotesize\ttfamily,numbers=none,keywordstyle=\color{blue}\bfseries,morekeywords={sudo}]{CMD} value, so that the only actual input is the value of \lstinline{PASSWORD}. In this restricted context, sudo has two input families: one where \lstinline[language=bash,basicstyle=\footnotesize\ttfamily,numbers=none,keywordstyle=\color{blue}\bfseries,morekeywords={sudo}]{PASSWORD} is a correct password and one where it is not, leading to two different effects on the environment (either \lstinline[language=bash,basicstyle=\footnotesize\ttfamily,numbers=none,keywordstyle=\color{blue}\bfseries,morekeywords={sudo}]{CMD} is executed as \lstinline[language=bash,basicstyle=\footnotesize\ttfamily,numbers=none,keywordstyle=\color{blue}\bfseries,morekeywords={sudo}]{alice} or an error message is printed). Second, let us assume that the effect of a program on its environment can be observed by recording the set of \emph{system calls} that it issues.
The rationale for this is that the PUT's interactions with the environment must be mediated by the operating system, which is achieved via system calls.

We then use the metamorphic oracle to detect the backdoor as follows.
First, record the system calls issued by sudo when fed an incorrect password value.
Then, fuzz sudo with only incorrect password values and compare the issued system calls with the recorded
ones. If a significant difference is spotted, then the metamorphic relation for the family of
incorrect passwords is violated and a potential backdoor is reported. This allows to detect the
\lstinline[language=C,basicstyle=\footnotesize\ttfamily,numbers=none,keywordstyle=\color{blue}\bfseries,morekeywords={sudo}]{"let_me_in"}
password, as the execution of
\lstinline[language=bash,basicstyle=\footnotesize\ttfamily,numbers=none,keywordstyle=\color{blue}\bfseries,morekeywords={sudo}]{CMD}
in a child process will trigger different system calls than printing an error message. Recall that modern fuzzers have mechanisms to quickly discover hard-coded ``magic'' values such as \lstinline[language=C,basicstyle=\footnotesize\ttfamily,numbers=none,keywordstyle=\color{blue}\bfseries,morekeywords={sudo}]{"let_me_in"}, as detailed in \Cref{subsubsec:magic-byte-fuzzing}.

In practice, all backdoors collected to build our \rosarum benchmark also result in divergent
system calls when triggered, because that is the only way for them to perform a meaningful task in
the PUT's environment, no matter how small it may appear.
As a consequence, all of them could possibly be detected by the metamorphic oracle described above.
Yet, in order to enable such a detection, the oracle should not only be used on a single input
family but on all input families deemed important enough to be searched for backdoors. However, most PUTs have numerous different input families. In the case of sudo, if we do not restrict the considered inputs to the password only, but also consider the impersonating and impersonated users, the command to be executed and the many flags that can be activated, we would end up with a combinatorial explosion of the number of families to individuate (manually) and then fuzz. To solve this issue, \rosa does not assume any prior knowledge of the PUT's input families, but instead relies on a heuristic method to automatically identify, for whatever input generated by a fuzzer, another input that should belong to the same family. The system calls issued when running the PUT with the two inputs are then compared to detect the presence of a backdoor.

\section{The \rosa approach}
\label{sec:approach}

\subsection{General overview}
\label{subsec:general-overview}

\begin{figure*}
    \centering

\begin{tikzpicture}[remember picture, node distance=0.2cm]

    \tikzstyle{node} = [
        text centered,
        font=\scriptsize,
    ]
    \tikzstyle{block} = [
        node,
rectangle,
        rounded corners,
        draw=black,
        fill=white,
        general shadow={fill=black, shadow yshift=-1pt},
    ]
    \tikzstyle{arrow} = [thick, ->, >=stealth]

    \newlength{\iconheight}
    \setlength{\iconheight}{0.6cm}

\node (phase-1) [
        block,
        font=\scriptsize,
        minimum width=1.2cm,
        label={[text depth=0pt] above:{\footnotesize\textbf{Phase 1}}},
        label={[text depth=0pt] below:{\footnotesize Representative inputs collection}}
    ] {
        \begin{tikzpicture}[node distance=1.5cm and 0.1cm]
\node (seed-put) [node, label=below:{\color{blue} Program}] {
                \includegraphics[height=\iconheight]{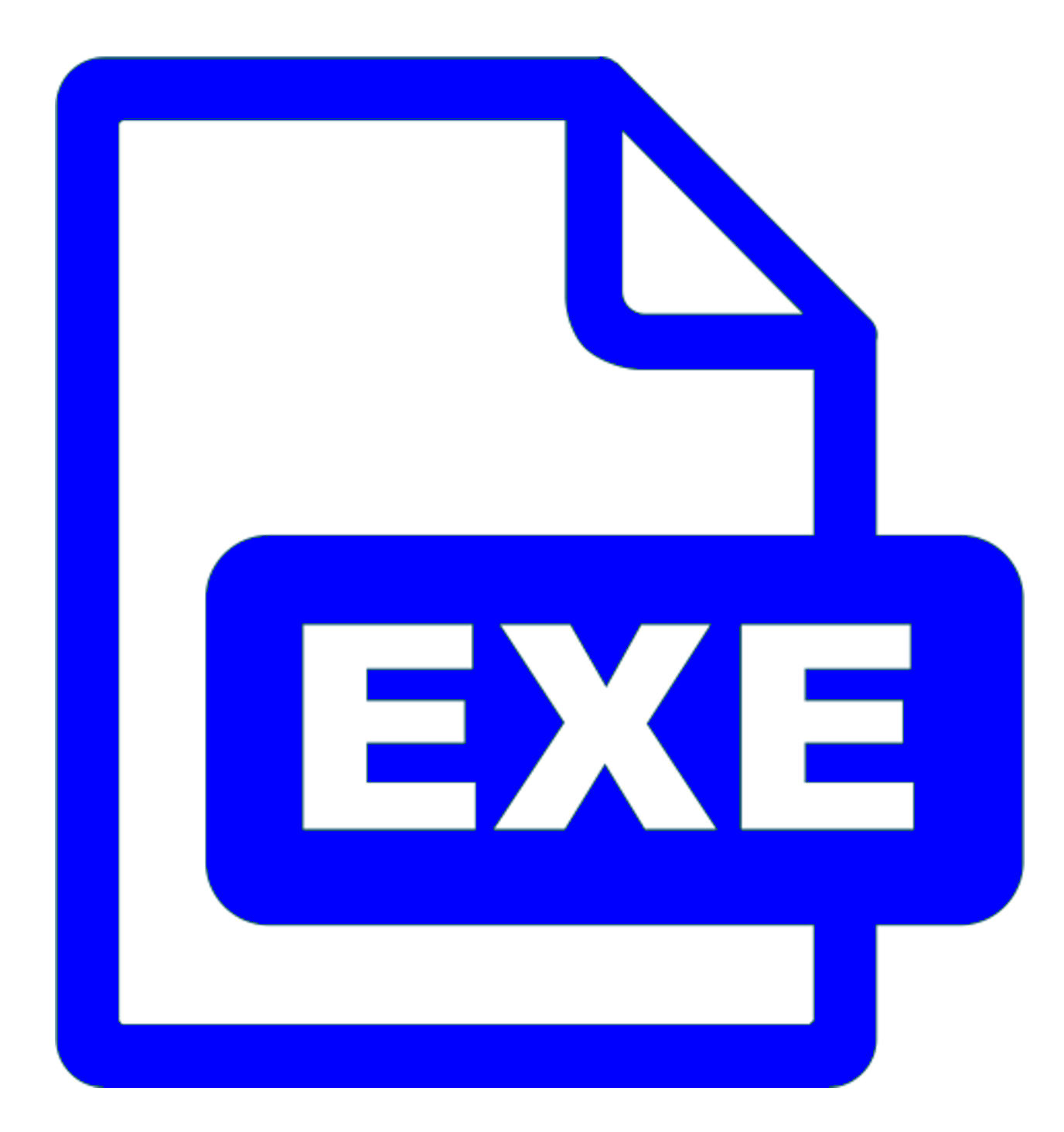}
            };
            \node (seed-fuzzer) [
                node, right=of seed-put, xshift=0.5cm, label=below:{\color{blue} Fuzzer}
            ] {
                \includegraphics[height=\iconheight]{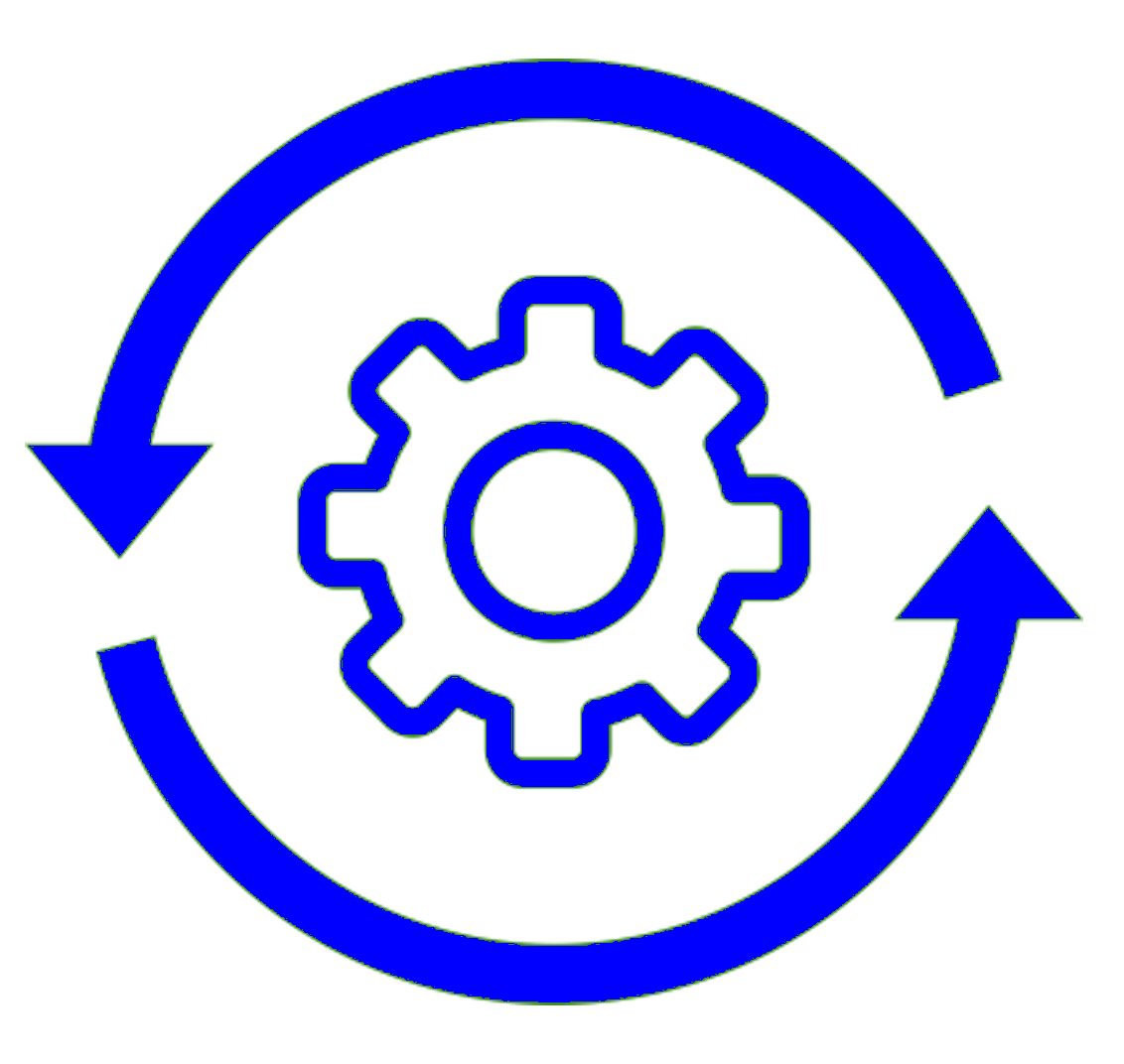}
            };
            \node (seed-database) [
                node,
                below=of seed-put,
                label={
                    [text width=2cm, text centered, name=seed-database-label]
                    below:Final input database
                }
            ] {
                \includegraphics[height=\iconheight]{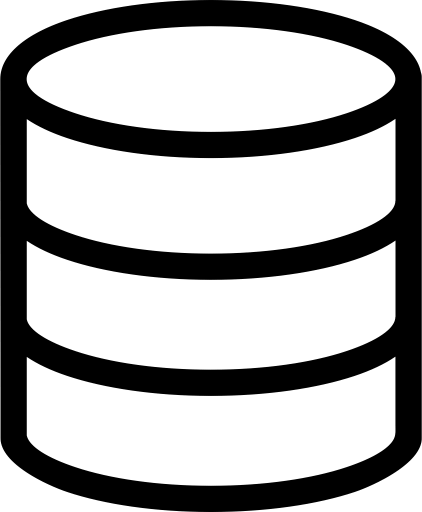}
            };

\node (seed-filter) [
                node,
                below=of seed-fuzzer,
                label={
                    [text width=2.1cm, text centered]
                    below:Filter inputs covering same edges
                },
            ] {
                \includegraphics[height=\iconheight]{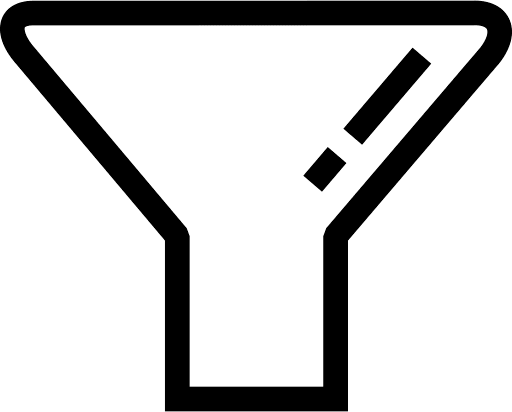}
            };

\draw [blue,arrow] ([xshift=-0.2cm] seed-put.east)
                -- ([xshift=0.2cm] seed-fuzzer.west);
            \draw [arrow] ([yshift=-0.4cm] seed-fuzzer.south)
                -- ([xshift=0.3cm] seed-database.north);
            \draw [arrow] ([xshift=-0.2cm]seed-database.east) -- ([xshift=0.2cm] seed-filter.west);
        \end{tikzpicture}
    };

\node (phase-2) [
        block,
        right=of phase-1,
        label={[text depth=0pt] above:{\footnotesize \textbf{Phase 2}}},
        label={[text depth=0pt] below:{\footnotesize Backdoor detection}}
    ] {
        \begin{tikzpicture}[node distance=1.5cm and 0.5cm]
\node (detection-put) [node, label=below:{\color{blue} Program}] {
                \includegraphics[height=\iconheight]{exe-file-icon}
            };
            \node (detection-fuzzer) [
                node,
                right=of detection-put,
                xshift=0.5cm,
                label=below:{\color{blue} Fuzzer},
            ] {
                \includegraphics[height=\iconheight]{fuzzer-icon}
            };
            \node (detection-generated-input) [
                node,
                right=of detection-fuzzer,
                xshift=0.6cm,
                label={
                    [text width=1.2cm, text centered, name=detection-generated-input-label]
                    below:Generated test input
                },
            ] {
                \includegraphics[height=\iconheight]{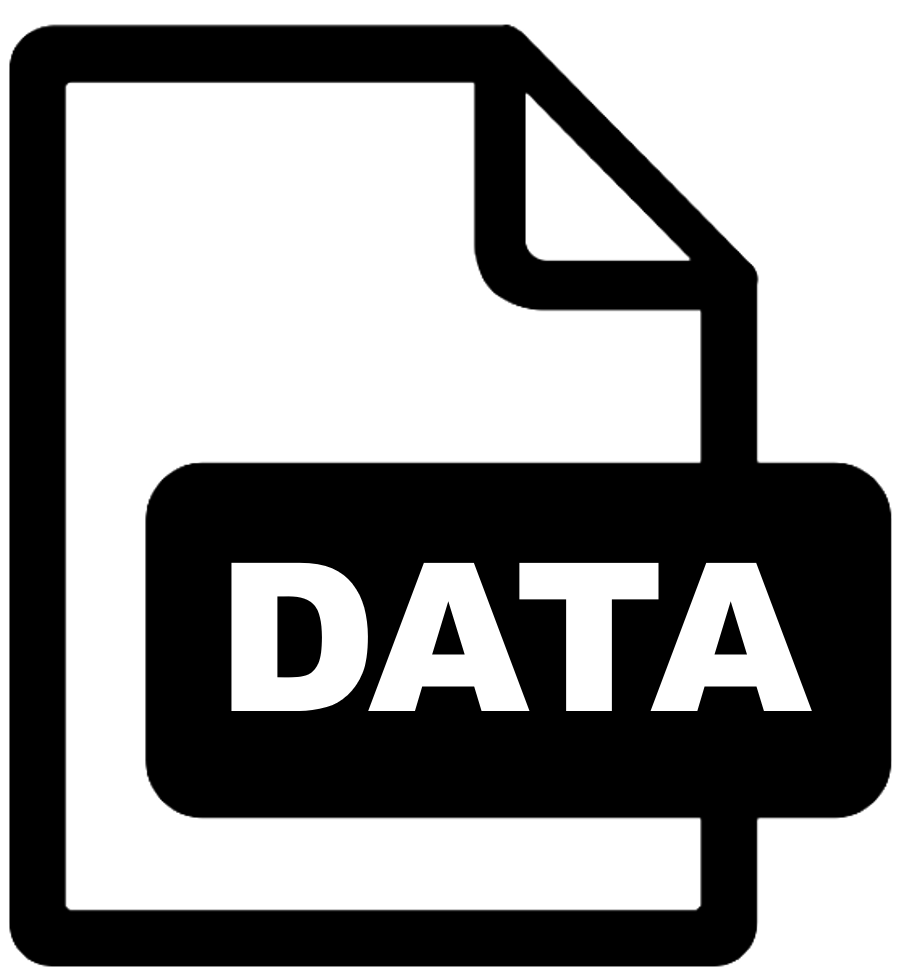}
            };
            \node (alt-put-top) [
                node,
                right=of detection-generated-input,
                xshift=0.6cm,
                label=below:{\color{blue} Program}
            ] {
                \includegraphics[height=\iconheight]{exe-file-icon}
            };
            \node (detection-input-syscalls) [
                node,
                right=of alt-put-top,
                xshift=0.2cm,
                label={
                    [text width=1.4cm, text centered]
                    below:Issued system calls
                },
            ] {
                \includegraphics[height=\iconheight]{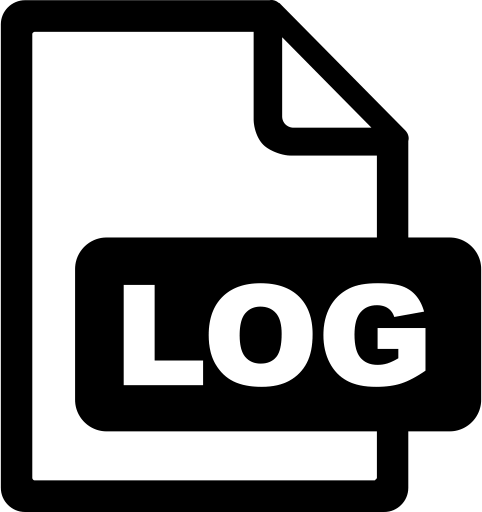}
            };

\node (detection-database) [
                node,
                below=of detection-fuzzer.south west,
                xshift=-0.5cm,
                label={
                    [text width=3cm, text centered]
                    below:Database of family-representative inputs
                },
            ] {
                \includegraphics[height=\iconheight]{database-icon}
            };
            \node (detection-family-input) [
                node,
                below=of detection-generated-input,
                label={
                    [text width=2.0cm, text centered]
                    below:Closest family-representative input
                },
            ] {
                \includegraphics[height=\iconheight]{data-file-icon}
            };
            \node (alt-put-bottom) [
                node,
                below=of alt-put-top,
                label=below:{\color{blue} Program}
            ] {
                \includegraphics[height=\iconheight]{exe-file-icon}
            };
            \node (detection-family-syscalls) [
                node,
                below=of detection-input-syscalls,
                label={
                    [text width=1.4cm, text centered]
                    below:Issued system calls
                },
            ] {
                \includegraphics[height=\iconheight]{log-file-icon}
            };

\node (oracle) [
                node,
                right=of detection-family-syscalls.north east,
                xshift=-0.3cm,
                yshift=0.5cm,
                text width=1.3cm
            ] {
                Dissimilarity?
            };

\draw [blue,arrow] (detection-put.east) -- (detection-fuzzer.west);
            \draw [arrow] (detection-fuzzer.east) -- (detection-generated-input.west);
            \draw [arrow] (detection-generated-input.east) -- (alt-put-top.west);
            \draw [arrow] (alt-put-top.east) -- (detection-input-syscalls.west);
            \draw [arrow] (detection-input-syscalls.east)
                -- ([xshift=0.9cm] detection-input-syscalls.east)
                -- (oracle.north);

\draw [dashed,arrow] (detection-database.east) -- (detection-family-input.west);
            \draw [arrow] (detection-family-input.east) -- (alt-put-bottom.west);
            \draw [arrow] (alt-put-bottom.east) -- (detection-family-syscalls.west);
            \draw [arrow] (detection-family-syscalls.east)
                -- ([xshift=0.9cm] detection-family-syscalls.east)
                -- (oracle.south);

\draw [dashed,arrow] (detection-generated-input-label.south west)
                -- node [midway, above, sloped, align=center,font=\tiny\linespread{0.8}] {{Input with} \\  {closest edge coverage?}} (detection-database.north east) ;
        \end{tikzpicture}
    };

\node (post) [
    block,
    font=\scriptsize,
    right=of phase-2,
    minimum width=1.4cm,
    label={[text depth=0pt] above:{\footnotesize \textbf{Post-processing}}},
    label={[text depth=0pt] below:{\footnotesize Deduplication and vetting}}
] {
    \begin{tikzpicture}[node distance=1.5cm and 0.1cm]
\node (report) [node,  label={[text width=1cm, text centered] below:{\color{red} Backdoor report}}] {
            \includegraphics[height=\iconheight]{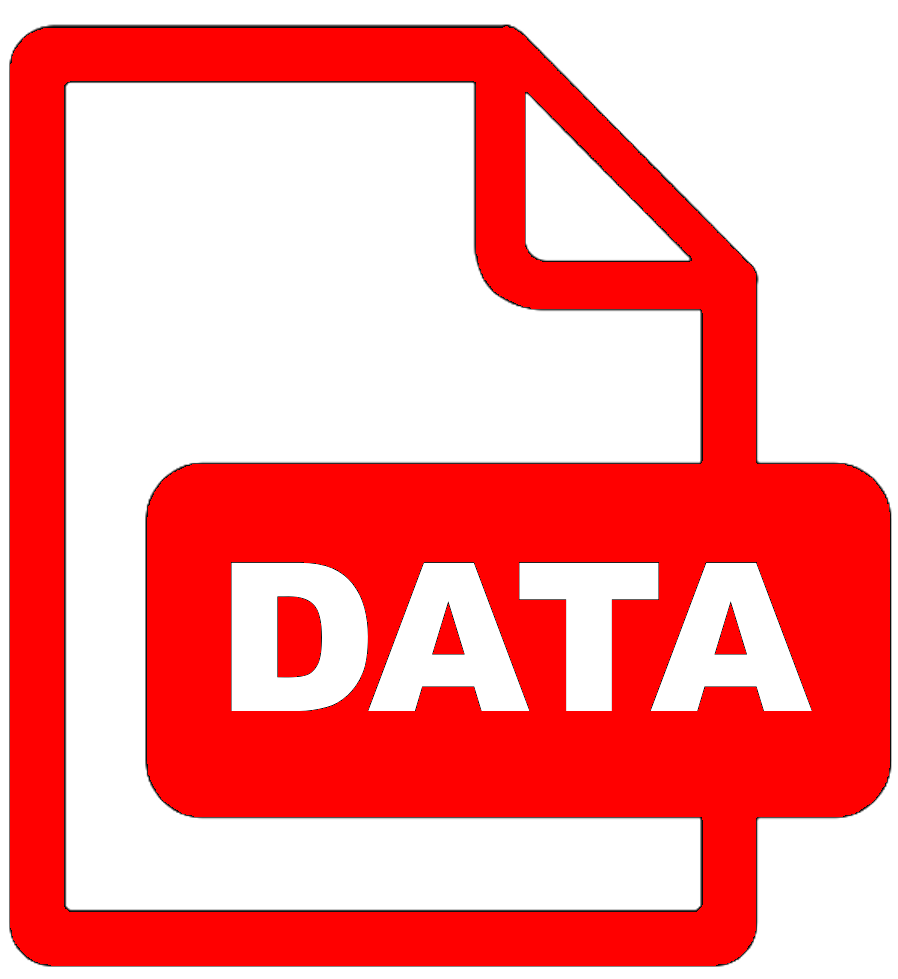}
        };
        \node (deduplication) [
            node, right=of report, xshift=0.5cm, label={[text width=1cm, text centered] below:{Filter duplicate reports}}]
         {
            \includegraphics[height=\iconheight]{filter-icon}
        };
\node (vetting) [
            node,
            below=of report,
            xshift=0.8cm,            
            label={
                [text width=2.5cm, text centered]
                below:Vetting of unique report by tooled expert
            },
        ] {
            \includegraphics[height=\iconheight]{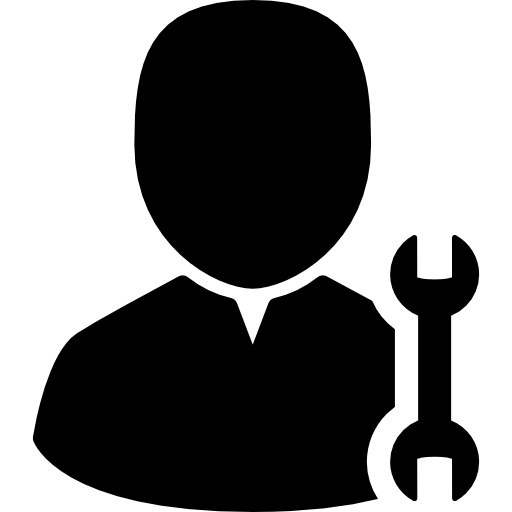}
        };

\draw [arrow] ([xshift=-0.3cm] report.east)
            -- ([xshift=0.3cm] deduplication.west);
        \draw [arrow] ([yshift=-1.0cm] deduplication.south)
            -- ([xshift=0.0cm] vetting.north);
    \end{tikzpicture}
};

\draw [arrow] ([xshift=-0.3cm] seed-filter.east) -- (detection-database.west);
    \draw [red, arrow] ([xshift=-0.3cm, yshift=0.3cm] oracle.east) -- node [near start, above, sloped, align=center,font=\scriptsize\linespread{0.8}] {{Yes}} ([xshift=0.3cm]report.west);

\end{tikzpicture}
     \vspace{-4mm}
    \caption{General overview of the \rosa approach to fuzzing-based backdoor detection.}
    \vspace{-3mm}
    \label{fig:rosa-process}
\end{figure*}

The \rosa approach to fuzzing-based backdoor detection unfolds in two successive phases, followed by a post-processing step, schematized in \Cref{fig:rosa-process}. (1) During the \emph{representative inputs collection} phase, we use a fuzzer on the PUT, in order to populate a database with a wide range of inputs that are deemed representative of its input families.
(2) Then, during the \emph{backdoor detection} phase, we fuzz the PUT for a much longer time. For every input generated by the fuzzer, we search the database of family-representative inputs, in order to heuristically identify another input that should belong to the same family as the generated input. The system calls issued when running the PUT with these two inputs are then compared. Following the principle of our metamorphic oracle, if a dissimilarity is spotted, a possible backdoor is signaled. The two inputs and the different system calls are then deduplicated and reported to the user for expert vetting.

In phase 1 (representative inputs collection), the challenge is to guide a fuzzer towards generating as many relevant family-representative inputs as possible. While inputs in the same input family produce a similar effect on the PUT environment, they can still trigger slightly different PUT behaviors. For example, wrong passwords in sudo could trigger different input sanitizing strategies, but should all eventually result in similar system calls. Consequently, we define a set of \emph{representative inputs of a family} as any set of inputs from the family, where each input triggers a different internal behavior, among those encompassed by the family.

In practice, the internal behavior of a program can be reasonably characterized by which edges of
its CFG are visited. A set of representative inputs then becomes any set of inputs from the
family, where each input in the set triggers the execution of a different combination of CFG edges.
In order to build our database of family-representative inputs, we thus need to identify and
combine sets of inputs from each family, where all inputs in a set cover different CFG edges. This
can be done by generating inputs with a fuzzer, identifying the corresponding family of each input,
and adding it to the database only if the corresponding family set in the database does not already
contain an input covering the same CFG edges. Yet, by the definition of input families, two inputs
covering exactly the same CFG edges also belong to the same family. Hence, two distinct family sets
in the database will never contain inputs covering exactly the same CFG edges and our procedure to
populate the database can be performed without prior knowledge of what the input families are. One
can indeed just fuzz the PUT and retain all of the
generated inputs that uncover \emph{new} CFG edge combinations in the database.  

During phase 2 (backdoor detection), the main challenge is to try and identify which inputs from
the representative inputs database are the most likely to belong to the same family as the inputs
generated by the fuzzer. We do this by searching the database for the input that covers the closest
combination of CFG edges, compared to each fuzzer-generated input. Here again, the rationale is
that an input family basically defines a class of slightly different internal PUT behaviors, each
best characterized by which CFG edges are visited.

\subsection{Phase 1: representative inputs collection}
\label{subsec:rosa-phase-i}

During phase 1, we fuzz the PUT and populate the representative input database with all of the generated inputs that have uncovered new CFG edges. In practice, we do this by taking advantage of the existing features of common graybox fuzzers that rely on edge coverage for guidance (like the many fuzzers based on AFL). After fuzzing, we collect the seed inputs produced by the fuzzer and filter out redundant ones, as some seeds may still cover the same combination of edges. 

The main stake of populating the representative inputs database in phase 1 is to properly sample the legitimate input families that will be met at phase 2. 
We call \emph{family subsampling} the situation where only some of these families would be discovered by the fuzzer in phase 1. In this case, no representative input is produced for the undiscovered families. These representative inputs would thus be missing when searching the database in phase 2. This will increase the risk of family misidentification, leading to false positives being reported, as the metamorphic oracle may mistakenly report some inputs as triggering a backdoor. 
We call \emph{backdoor contamination} the situation where the fuzzer ends up triggering a backdoor in phase 1, despite the fact that backdoors are designed to be activated by a small fraction of inputs, and are thus unlikely to trigger in phase 1. In case of backdoor contamination, backdoor-triggering inputs may end up recorded incorrectly as representative inputs of legitimate families, and used for family identification in phase 2. This will increase the risk of the metamorphic oracle mistakenly classifying some inputs as legitimate. Yet, backdoors can usually be triggered by inputs belonging to many different input families (for example, the sudo backdoor from \Cref{lst:sudo_backdoor} can be triggered with many different combinations of command-line flags) and, as backdoor contamination is unlikely to occur for \emph{all} these families all at once, the backdoor would still be detectable with inputs from the uncontaminated families.

In practice, the levels of family subsampling and backdoor contamination can be controlled by adjusting the duration of the phase-1 fuzzing campaign. The longer the campaign, the lower the risk of family subsampling and the higher the risk of backdoor contamination. In \Cref{sec:evaluation}, we perform a parameter sweep study for campaign durations between 30 seconds and 20 minutes, showing that all durations provide acceptable results on our \rosarum backdoor benchmark.

\subsection{Phase 2: backdoor detection}
\label{subsec:rosa-phase-ii}

During phase 2, we fuzz the PUT again with a graybox fuzzer, but this time for as long as possible (i.e., as available resources permit, like in traditional fuzzing). For each input generated by the fuzzer, we retrieve the input that covers the closest combination of CFG edges from the representative inputs database. We run the PUT with these two inputs, and signal a backdoor if they do not trigger similar system calls. 

\textbf{Example.}
Let us imagine that the graybox fuzzer has generated an input called \textbf{\small Input \#1}. We run the PUT on it and record which CFG edges are covered and which types of system calls are used (among all those allowed by the operating system API, like \lstinline{read}, \lstinline{kill},
\lstinline{open} and others in Linux). The results can be stored in vectors as follows (where \textbf{\small Input \#1} covers the four edges of the PUT and uses all three available system call types, but \lstinline{kill}):

\begin{center}
\begin{tcolorbox}[width=0.34\textwidth,boxrule=0.5pt,colback=white]
\vspace{-2mm}
\scriptsize
\centering
\underline{\textbf{Input \#1}} \begin{tabular}{c|c|c|c|c|}
\cline{2-5}
\multirow{2}{*}{\color{blue}CFG edges} &  1 & 2 & 3 & 4 \\
\cline{2-5}
 & \ding{51}  & \ding{51} & \ding{51}  & \ding{51} \\
\cline{2-5}
\end{tabular}\\
~

\begin{tabular}{c|c|c|c|}
\cline{2-4}
\multirow{2}{*}{\color{red}System calls} & \texttt{read} & \texttt{kill} & \texttt{open}  \\
\cline{2-4}
 & \ding{51}  & \ding{55} & \ding{51}  \\
\cline{2-4}
\end{tabular}
\vspace{-2mm}
\end{tcolorbox}
\end{center}

\noindent
Let us now retrieve the input that covers the closest combination of CFG edges from the representative inputs database. Let us imagine that this database contains two inputs (\textbf{\small Input \#A} and \textbf{\small Input \#B}), whose covered edges and used system call types are as follows:

\begin{center}
\begin{tcolorbox}[width=0.4\textwidth,colback=white,colframe=blue,boxrule=0.5pt,]
\vspace{-2mm}
\centering
\footnotesize
\textbf{\color{blue}\ttfamily \scriptsize Representative Inputs Database}

\begin{tcolorbox}[width=0.98\textwidth,boxrule=0.5pt,colback=white]
\vspace{-2mm}
\scriptsize
\centering
\underline{\textbf{Input \#A}} \begin{tabular}{c|c|c|c|c|}
\cline{2-5}
\multirow{2}{*}{\color{blue}CFG edges} &  1 & 2 & 3 & 4 \\
\cline{2-5}
 & \ding{51}  & \ding{55} & \ding{51}  & \ding{55} \\
\cline{2-5}
\end{tabular}\\
~

\begin{tabular}{c|c|c|c|}
\cline{2-4}
\multirow{2}{*}{\color{red}System calls} & \texttt{read} & \texttt{kill} & \texttt{open}  \\
\cline{2-4}
 & \ding{55}  & \ding{55} & \ding{51}  \\
\cline{2-4}
\end{tabular}
\vspace{-2mm}
\end{tcolorbox}

\begin{tcolorbox}[width=0.98\textwidth,boxrule=0.5pt,colback=white]
\vspace{-2mm}
\scriptsize
\centering
\underline{\textbf{Input \#B}} \begin{tabular}{c|c|c|c|c|}
\cline{2-5}
\multirow{2}{*}{\color{blue}CFG edges} &  1 & 2 & 3 & 4 \\
\cline{2-5}
 & \ding{55}  & \ding{51} & \ding{55}  & \ding{55} \\
\cline{2-5}
\end{tabular}\\
~

\begin{tabular}{c|c|c|c|}
\cline{2-4}
\multirow{2}{*}{\color{red}System calls} & \texttt{read} & \texttt{kill} & \texttt{open}  \\
\cline{2-4}
 & \ding{55}  & \ding{55} & \ding{55}  \\
\cline{2-4}
\end{tabular}
\vspace{-2mm}
\end{tcolorbox}
\vspace{-3mm}
\end{tcolorbox}
\end{center}

\noindent
To establish which of \textbf{\small Input \#A} and \textbf{\small Input \#B} has the closest
combination of CFG edges compared to \textbf{\small Input \#1}, we compute the Hamming
distance~\cite{hamming-1980} between the CFG edge vectors and consider the one for which the distance is the smallest. In this case, \textbf{\small Input \#A} is retrieved, as its coverage w.r.t.~\textbf{\small Input \#1} differs only by two edges, against three for \textbf{\small Input \#B}:

\begin{center}
\scriptsize
\begin{tabular}{|c|c|c|c|c|c|c|c|c|c|c|}
\cline{2-5}\cline{8-11}
\multicolumn{1}{c|}{} &  1 & 2 & 3 & 4 & \multicolumn{1}{c}{} & \multicolumn{1}{c|}{} &  1 & 2 & 3 & 4 \\
\cline{1-5}\cline{7-11}
Input \#1 & \ding{51}  & \color{blue} \ding{51} & \ding{51}  & \color{blue} \ding{51} & & Input \#1 & \color{blue} \ding{51}  &  \ding{51} &  \color{blue}  \ding{51}  & \color{blue} \ding{51} \\
Input \#A & \ding{51}  & \color{blue} \ding{55} & \ding{51}  & \color{blue} \ding{55} & & Input \#B & \color{blue} \ding{55}  & \ding{51} &  \color{blue}  \ding{55}  & \color{blue} \ding{55} \\
\cline{1-5}\cline{7-11}
\multicolumn{5}{|c|}{ \color{blue} Hamming distance for edges = 2} & & \multicolumn{5}{c|}{ \color{blue} Hamming distance for edges = 3}  \\
\cline{1-5}\cline{7-11}
\end{tabular}
\end{center}

\noindent
Finally, system call comparison is done by computing the Hamming distance between the system call type vectors of \textbf{\small Input \#1} and the retrieved input \textbf{\small Input \#A}:
\begin{center}
\scriptsize
\begin{tabular}{|c|c|c|c|}
\cline{2-4}
\multicolumn{1}{c|}{} & \texttt{read} & \texttt{kill} & \texttt{open}  \\
\hline
 Input \#1  & \color{red} \ding{51}  & \ding{55} & \ding{51}   \\
Input \#A & \color{red} \ding{55}  & \ding{55} & \ding{51}  \\
\hline
\multicolumn{4}{|c|}{ \color{red} Hamming distance for system calls = 1}  \\
\hline
\end{tabular}
\end{center}

\noindent
We report a backdoor as soon as this distance is non-zero. In this case, a backdoor is reported
because \textbf{\small Input \#1} uses a \lstinline{read} system call and \textbf{\small Input \#A}
does not. Note that if, during database retrieval, the Hamming distance between CFG edge vectors had been the same for \textbf{\small Input \#A} and \textbf{\small Input \#B}, we would have reported a backdoor only if the Hamming distance between the system call type vectors had been non-zero for \emph{both} \textbf{\small Input \#A} and \textbf{\small Input \#B}.

\subsection{Post-processing: deduplication and vetting}
\label{subsec:rosa-post}

Fuzzers are stochastic and thus subject to repeatedly triggering the same issues in the PUT. As a consequence, they often use deduplication techniques, to try and avoid polluting their output with duplicated reports (see Donaldson et al. \cite{donaldson-2021} for an example). \rosa follows a similar approach and reports a backdoor only if it had not previously reported another backdoor involving the same system call difference with the same family-representative input, as these two reports are likely to describe the same backdoor. Our experiments show that this heuristic enables reducing by one order of magnitude on average the number of reports produced by \rosa, without impeding its ability to find backdoors.

As family identification is made heuristically, \rosa is capable of returning false positives. Each
backdoor report (of the form ``\textbf{\small Input \#1} is suspicious because it uses a
\lstinline{read} system call and \textbf{\small Input \#A} does not'') must thus be manually vetted by an expert (an
understanding of the expected and actual program behaviors is indeed needed here, for which there is currently no fully-automatic solution). Yet, vetting can still occur in a pretty systematized and semi-automated way:
    \begin{enumerate}
        \item Run the PUT under a process tracing tool (like \lstinline{strace}~\cite{strace}) with both reported 
            inputs, filtering out all system calls except the ones listed in the report;
        \item Compare the filtered traces to determine whether one represents a privilege escalation within the program or an undue access to the underlying system.
    \end{enumerate}

\textbf{Example.} 
Let us illustrate how vetting works on a real (positive) report returned by \rosa on the authentic ProFTPD backdoor~\cite{schuster-2013}. Running the suspicious input under \lstinline{strace} and keeping only the divergent system calls produces the following pattern:
\lstinline{clone3(...)}, \lstinline{setuid(0)}, \lstinline{setgid(0)}, \lstinline{execve("/bin/sh", ...)},
which is a transparent attempt at spawning a root shell (not an expected behavior for ProFTPD). The suspicious input itself, 14-lines long, contains the surprising \lstinline{HELP ACIDBITCHEZ} command, which can be identified as part of the key input value of a backdoor.

\section{Implementation}
\label{sec:implementation}
  
We have implemented the \rosa tool by relying on AFL++~\cite{fioraldi-2020} (version
\lstinline{++4.20c}) for the fuzzing campaigns of both phases 1 and 2. The \rosa tool needs to be
provided with the PUT and a corpus of initial seed inputs to fuzz it. Its main parameters are the
respective lengths of the two fuzzing campaigns. At the end, the tool returns a list of PUT inputs
that trigger potential backdoors, similar to how vanilla AFL++ returns a list of crash-triggering
inputs. These inputs can then be vetted as discussed above.

As backdoor detection must often be performed on PUTs that come in binary-only form, we use AFL++ in binary-only mode~\cite{qemu-afl}, with QEMU~\cite{bellard-2005} as backend emulator. Otherwise, we configure AFL++ by following the best practices described in its documentation. This notably means using six synchronized instances of the fuzzer running in parallel, with different seed prioritization and mutation strategies. Half of the instances only instrument the PUT itself, while the other half also instrument the called external library functions, enabling backdoor detection in dynamically loaded libraries. All instances leverage the built-in AFL++ mechanisms to deal efficiently with magic byte comparisons \cite{laf-intel-2016, aschermann-2019}, as these are often used as backdoor triggers. It should be noted that support for the configuration of AFL++ that we have just described is currently limited to fuzzing x86/x64 binaries on Linux. Our tool inherits thus this limitation.

Finally, during the fuzzing campaigns of phases 1 and 2, our tool records on the fly which CFG edges and system calls are triggered by the generated inputs, for later use according to the \rosa approach. This is implemented through light modifications to the AFL++ and QEMU code. For recording edge coverage data, we rely on the existing edge coverage measurement mechanisms of the fuzzer. For recording system call coverage data, we implement similar mechanisms to those used for edge coverage measurement, but we track system call instructions instead of jump instructions between basic blocks.

\section{Experimental evaluation}
\label{sec:evaluation}

\subsection{General overview}

We aim at answering the following research questions:
\begin{description}
    \item[\textbf{RQ1}] Can \rosa detect backdoors in enough diverse contexts, with enough robustness, speed and automation, to make it usable and useful in the wild?
    \item[\textbf{RQ2}] How does \rosa compare to state-of-the-art backdoor detection tools, in terms of robustness, speed and automation?
\end{description}

\noindent
To answer these, we need to run \rosa and competing tools over samples of backdoors in real programs. Yet, there is no existing off-the-shelf backdoor dataset that could be leveraged to do so. Indeed, previous papers introducing program analysis tools for backdoor detection~\cite{schuster-2013, shoshitaishvili-2015, thomas-stringer-2017, thomas-humidify-2017} use different small backdoor datasets for evaluation purposes. In addition, these papers date back 7--11 years, so that some samples have been lost, while those that are still available can be undocumented binary firmware, running only on old IoT devices.
As a preliminary step to our evaluation, we have thus \textit{assembled a novel backdoor benchmark dataset, called \rosarum}.

\subsection{Constructing the \rosarum benchmark}
\label{subsec:eval-benchmark}

\begin{table*}\caption{List of the \BackdoorsAuthentic authentic and \BackdoorsSynthetic synthetic backdoors that form our new \textbf{\rosarum benchmark} for backdoor detector evaluation.
  }
  \label{tab:benchmark}
  \centering

\begin{tabular}{| l | l | r | P{1.75cm} | l |}
    \hline

    \multicolumn{3}{|c|}{\textbf{Program}}
        & \multicolumn{2}{c|}{\textbf{Backdoor}} \\ 
    
    \multicolumn{1}{|c}{\textbf{Name}}
        & \multicolumn{1}{c}{\textbf{Type}}
        & \multicolumn{1}{c|}{\textbf{Binary size}}
        & \multicolumn{1}{c}{\textbf{Origin}} 
        & \multicolumn{1}{c|}{\textbf{Description}} \\\hline\hline

    \rowcolor{blue!10}
    \multicolumn{5}{|c|}{\textbf{Authentic backdoors}} \\\hline

    Belkin / httpd 
        & Router HTTP server
        & \num{2.6} MiB & \multirow{4}{1.75cm}{Router manufacturer }
        & HTTP request with secret URL value leads to web shell \cite{toterhi-2018}
        \\
        \cline{1-3}
        \cline{5-5}
        
    D-Link / thttpd 
        & Router HTTP server
        & \num{7.2} MiB
        & 
        & HTTP request with secret field value bypasses authentication \cite{lee-2013}
        \\
        \cline{1-3}
        \cline{5-5}

    Linksys / scfgmgr 
        & Router TCP server
        & \num{2.5} MiB
        & 
        & Packet with specific payload enables memory read/write \cite{benoist-vanderbeken-2015}
        \\
        \cline{1-3}
        \cline{5-5}

    Tenda / goahead 
        & Router HTTP server
        & \num{2.9} MiB
        & 
        & Packet with specific payload enables command execution \cite{craig-2013}
        \\\hline

    PHP
        & HTTP server
        & \num{80.6} MiB
        &  \multirow{3}{1.75cm}{Supply-chain attack } 
        & HTTP request with secret field value enables command execution \cite{ganz-2023}
        \\
        \cline{1-3}
        \cline{5-5}

    ProFTPD 
        & FTP server
        & \num{3.3} MiB
        & 
        & Secret FTP command leads to root shell \cite{schuster-2013}
        \\
        \cline{1-3}
        \cline{5-5}

    vsFTPd 
        & FTP server
        & \num{2.9} MiB & 
        & FTP usernames containing \texttt{":)"} lead to root shell \cite{evans-2011}
        \\\hline\hline

    \rowcolor{red!10}
    \multicolumn{5}{|c|}{\textbf{Synthetic backdoors}} \\\hline

    sudo
        & Unix utility
        & \num{8.4} MiB
        &  \multirow{1}{*}{Paper example}
        & Hardcoded credentials (see \Cref{lst:sudo_backdoor}) \\\hline

    libpng
        & Image library
        & \num{7.0} MiB
        & \multirow{9}{1.75cm}{Manual injection in the MAGMA~\cite{hazimeh-2020} fuzzing benchmark
        }
        & Secret image metadata values enables command execution \\
        \cline{1-3}
        \cline{5-5}

    libsndfile
        & Sound library
        & \num{6.6} MiB 
        & {}
        & Secret sound file metadata value triggers home directory encryption  \\
        \cline{1-3}
        \cline{5-5}

    libtiff
        & Image library
        & \num{10} MiB
        & {}
        & Secret image metadata value enables command execution \\
        \cline{1-3}
        \cline{5-5}

    libxml2
        & XML library
        & \num{8.2} MiB
        & {}
        & Secret XML node format enables command execution \\
        \cline{1-3}
        \cline{5-5}

    Lua
        & Language interpreter
        & \num{3.7} MiB
        & {}
        & Specific string values in script enables reading from filesystem \\
        \cline{1-3}
        \cline{5-5}
        
    OpenSSL / bignum
        & Crypto library
        & \num{12.2} MiB
        & {}
        & Secret bignum exponentiation string enables command execution \\
        \cline{1-3}
        \cline{5-5}

    PHP / unserialize
        & Language interpreter
        & \num{30.2} MiB
        & {}
        & Specific string values in serialized object enables PHP code execution \\
        \cline{1-3}
        \cline{5-5}

    Poppler
        & PDF renderer
        & \num{39.4} MiB
        & {}
        & Secret character in PDF comment enables command execution \\
        \cline{1-3}
        \cline{5-5}

    SQLite3
        & Database system
        & \num{6.4} MiB
        & {}
        & Secret SQL keyword enables removal of home directory \\\hline

\end{tabular}
 \end{table*}

\subsubsection{Collecting and porting authentic backdoors}
As a first step to populate \rosarum, we have collected authentic backdoors from three types of sources. First, we have looked into all state-of-the-art papers~\cite{thomas-2018, ganz-2023,  schuster-2013, shoshitaishvili-2015, thomas-stringer-2017, thomas-humidify-2017} for alive references to backdoor samples. When necessary, we have also contacted their authors to verify the availability of the backdoors mentioned in the papers. Second, we have searched public vulnerability and exploit databases for sufficiently documented reports that describe genuine code-level backdoors. Finally, we have searched general and IT news reports (gray literature) for references to code-level backdoors, and looked for the corresponding backdoor samples. In total, we have performed an in-depth sample search and analysis for 15 backdoor reports, including 11 from the state of the art, 1 from public databases and 3 from gray literature. Samples could not be obtained for 4 reports from the state of the art. As the \rosa tool only supports x86/x64 Linux binaries, we had to make sure that collected backdoors work on such a platform. While 5 samples already supported it natively, the 6 remaining ones did not. We invested significant effort in porting them. The RaySharp and QSee backdoors from the state of the art~\cite{thomas-stringer-2017} are the only ones that could not be ported, due to unresolved dependencies on libraries and IoT peripherals.

\subsubsection{Seeding synthetic backdoors in a fuzzing benchmark}
\label{subsubsec:synthetic-backdoors}
To further enrich \rosarum with more diverse programs and backdoors, we have followed an approach
used in the state of the art~\cite{schuster-2013}, where students were asked\footnote{Automatic injection of non-trivial backdoors in random code is a hard problem, and there exists no automatic backdoor injection tool to date.} to inject {synthetic
backdoors} into the ProFTPD program. We have hence asked a researcher \emph{not involved in the
development of \rosa} (in the style of clean-room design) to seed synthetic backdoors into the
programs of MAGMA~\cite{hazimeh-2020}, a recent fuzzing benchmark. These programs are ``open-source libraries with widespread usage and a long
history of security-critical bugs''~\cite{hazimeh-2020}. They are deemed challenging for modern fuzzers, so they are used to evaluate their vulnerability detection capabilities. The \rosa developers performed blind detection campaigns over these programs, with no knowledge of the backdoors' inner workings.

\subsubsection{Preparing the backdoors for benchmarking}
For every authentic or synthetic backdoor in \rosarum, we include the safe source of the program and two patches, a \emph{backdoor patch} and a \emph{ground-truth patch}. The backdoor patch simply injects the backdoor into the program source, while the ground-truth one inserts backdoor detection markers instead. These markers print a predefined string to denote the successful triggering of the backdoor, enabling one to verify if suspicious inputs reported by detection tools are true/false positives/negatives. Once a patch possibly applied, the source can be compiled into a x86/x64 Linux binary using the provided Makefile.

\subsubsection{Description of the final benchmark}
In total, \rosarum contains \BackdoorsTotal backdoors (\BackdoorsAuthentic authentic + \BackdoorsSynthetic synthetic, i.e. the largest benchmark ever used to evaluate backdoor detection techniques, as per the state of the art), detailed in \Cref{tab:benchmark}. The sudo backdoor of \Cref{lst:sudo_backdoor}, used earlier to illustrate the \rosa approach, is also included. On the other hand, the authentic Trendnet backdoor described in the state of the art~\cite{thomas-stringer-2017} is not included, as our analysis revealed that it was in fact a false positive, due to inoffensive hard-coded credentials. We have also excluded the recent authentic xz~\cite{CVE-2024-3094} backdoor, as our analysis revealed that triggering the backdoor requires passing a signature check with a hard-coded public key, so that only an attacker knowing the private key can do it. This mechanism prevents triggering the backdoor with a fuzzer or other dynamic analyses in reasonable time, as they need to brute-force the cryptography\footnote{The xz backdoor is particularly intricate and adversarial. It also includes dynamic code modifications that are hard to handle for static analyses.}. We elaborate further on this in \Cref{subsec:eval-rosa}, together with other limitations of \rosa.

\subsection{RQ1: usability and usefulness of \rosa}
\label{subsec:eval-rosa}

\subsubsection{Experimental protocol}
\label{subsubsec:experimental-protocol}
We design our experimental protocol by carefully adapting the best practices for fuzzing evaluation~\cite{klees-2018,schloegel-2024} to the backdoor case, 
for which no baseline exists. 
Specifically, we perform 10 independent 8-hour runs of \rosa for each backdoor in \rosarum. We
allocate 8 CPU cores and 16 GiB of RAM to each run, on a dedicated Intel\textregistered\
Xeon\textregistered\ Silver 4241 2.20\,GHz server. For the MAGMA~\cite{hazimeh-2020} programs,
we use the initial seeds provided by the benchmark; we use the standard seeds from the AFL++ documentation
  for HTTP\footnote{\url{https://securitylab.github.com/research/fuzzing-apache-1}} and
  FTP\footnote{\url{https://securitylab.github.com/resources/fuzzing-sockets-FTP}} servers; we use
  a single seed containing the string \lstinline{"test"} for the remaining programs.

We evaluate two different aspects of \rosa. First, we \textbf{measure the robustness and speed of \rosa}, in diverse contexts and at scale. 
To do so, we measure, for each \rosarum backdoor, (1) the proportion of \rosa runs that fail to find the backdoor before timing out, and (2) the minimum, average, and maximum time (over the 10 runs) needed by \rosa to find a first input that triggers the backdoor. 
Second, we \textbf{measure the level of automation of \rosa}, i.e. we evaluate how much additional
manual work is required by \rosa, compared to using a classical fuzzer like AFL++ for finding
crashes. The main overhead of \rosa comes from its ability to produce false positives, while crash
reports are usually legitimate. An expert must thus inspect all of the reported inputs one by one,
until they discover the backdoor or discard all of the inputs as false positives, following the
semi-automated procedure described at \Cref{subsec:rosa-post}. We estimate this manual expert effort by reporting, for each \rosarum backdoor, the minimum, average and maximum number of inputs that an expert should draw at random, either to have a 95\% probability of coming across an input that triggers the backdoor (for the runs that succeed in finding it) or to establish that no input triggers it (for the runs that fail). We also report on the time needed to vet an input with our semi-automated procedure.

\begin{table*}
  \caption{\textbf{Backdoor detection results} of our \rosa tool and the competing \textsc{Stringer}
    tool~\cite{thomas-stringer-2017}, on the \rosarum benchmark.\\
    Two evaluation goals are depicted: (1) robustness (whether a backdoor is found) + speed
    (how long it takes),\\
    and (2) automation level (number of inputs to vet manually,
    against total number of AFL++ seeds); see \Cref{subsec:eval-rosa} for
    details.\\
  }
  \label{tab:results-8h}
  \centering

\begin{tabular}{| l || c | c c c | c | c c c || c  | c |}
    \hline

    \multicolumn{1}{|c||}{\multirow{4}{*}{\textbf{Backdoor}}}
        & \multicolumn{8}{c ||}{\textbf{\rosa} --- (10 runs \texttimes~8 hours) / backdoor --- \BestPhaseOneDuration~of fuzzing for phase 1}
        & \multicolumn{2}{c |}{\textbf{\textsc{Stringer}}}\\
        & \multicolumn{4}{c |}{\textbf{Robustness + speed}}
        & \multicolumn{4}{c ||}{\textbf{Automation level}}
        & \multicolumn{1}{c|}{\multirow{3}{1.25cm}{\centering \textit{\textbf{Backdoor detection time}}}} 
        & \multicolumn{1}{c|}{\multirow{3}{1.25cm}{\centering \textit{\textbf{Manually inspected strings}}}}  \\
        & \multicolumn{1}{c|}{\multirow{2}{1cm}{\centering\textit{\textbf{Failed runs}}}} 
        & \multicolumn{3}{c|}{\textit{\textbf{Time to first backdoor input}}}
        & \multicolumn{1}{c|}{{\centering\textit{\textbf{Baseline}}}} 
        & \multicolumn{3}{c||}{\textit{\textbf{Manually inspected inputs}}}
        &
        & \\

    {}
        &
        & {\scriptsize \textbf{Min.}}
        & {\scriptsize \textbf{Avg.}}
        & {\scriptsize \textbf{Max.}}
        & {\scriptsize \textbf{Avg. seeds}}
        & {\scriptsize \textbf{Min.}}
        & {\scriptsize \textbf{Avg.}}
        & {\scriptsize \textbf{Max.}}
        &
        & \\\hline\hline

    \rowcolor{blue!10}
    \multicolumn{11}{|c|}{\textbf{Authentic backdoors}} \\\hline

    Belkin / httpd
        & 10 / 10
        & Timeout
        & Timeout
        & Timeout
        & 2773
        & 2
        & 4
        & 6
        & Not found
        & \textbf{0} \\

    {\scriptsize $\mathbf{+}$ \textit{with specialized seeds}*}
        & \textbf{3 / 10}
        & \textbf{17m40s}
        & \textbf{3h49m29s}
        & Timeout
        & 2781
        & 4
        & 5
        & 7 
        & Not found
        & \textbf{0} \\\hline

    D-Link / thttpd
        & \textbf{0 / 10}
        & \textbf{2m07s}
        & \textbf{15m00s}
        & \textbf{43m42s}
        & 3648
        & \textbf{7}
        & \textbf{9}
        & \textbf{12}
        & Not found
        & 113 \\\hline

    Linksys / scfgmgr
        & \textbf{0 / 10}
        & \textbf{1m05s}
        & \textbf{1m29s}
        & \textbf{1m55s}
        & 251
        & 1
        & 1
        & 1
        & Not found
        & \textbf{0} \\\hline

    Tenda / goahead
        & \textbf{0 / 10}
        & \textbf{1m28s}
        & \textbf{3m34s}
        & \textbf{8m10s}
        & 535
        & \textbf{1}
        & \textbf{2}
        & \textbf{2} 
        & Not found
        & 290 \\\hline

    PHP
        & 1 / 10
        & 24m30s
        & 2h03m44s
        & Timeout
        & 11631
        & \textbf{4}
        & \textbf{8}
        & \textbf{16}
        & \textbf{6m}
        & 573 \\\hline

    ProFTPD
        & 4 / 10
        & 4m03s
        & 3h37m32s
        & Timeout
        & 2995
        & \textbf{5}
        & \textbf{8}
        & \textbf{11}
        & \textbf{7s}
        & 314 \\\hline

    vsFTPd
        & \textbf{0 / 10}
        & \textbf{3m04s}
        & \textbf{5m41s}
        & \textbf{11m03s}
        & 1890
        & \textbf{3}
        & \textbf{4}
        & \textbf{4}
        & Not found
        & 117 \\\hline\hline

    \rowcolor{red!10}
    \multicolumn{11}{|c|}{ \textbf{Synthetic backdoors}} \\\hline

    sudo
        & \textbf{0 / 10}
        & \textbf{5m47s}
        & \textbf{8m05s}
        & \textbf{11m46s}
        & 167
        & \textbf{1}
        & \textbf{1}
        & \textbf{1} 
        & Not found
        & 137 \\\hline

    libpng
        & 2 / 10
        & 13m47s
        & 2h24m46s
        & Timeout
        & 4202
        & \textbf{1}
        & \textbf{2}
        & \textbf{2}
        & \textbf{4s}
        & 9 \\\hline

    libsndfile
        & 3 / 10
        & 2h21m08s
        & 5h04m46s
        & Timeout
        & 10376
        & 9
        & 12
        & 13
        & \textbf{5s}
        & \textbf{8} \\\hline

    libtiff
        & \textbf{0 / 10}
        & \textbf{5m08s}
        & \textbf{12m15s}
        & \textbf{25m10s}
        & 9566
        & \textbf{1}
        & \textbf{3}
        & \textbf{5}
        & Not found
        & 31 \\\hline

    libxml2
        & \textbf{0 / 10}
        & \textbf{8m17s}
        & \textbf{27m14s}
        & \textbf{1h09m06s}
        & 12104
        & \textbf{9}
        & \textbf{14}
        & \textbf{20}
        & Not found
        & 1208 \\\hline

    Lua
        & \textbf{1 / 10}
        & \textbf{50m34s}
        & \textbf{4h07m41s}
        & Timeout
        & 6653
        & \textbf{6}
        & \textbf{12}
        & \textbf{17}
        & Not found
        & 36 \\\hline

    OpenSSL / bignum
        & \textbf{0 / 10}
        & \textbf{9m53s}
        & \textbf{22m00s}
        & \textbf{39m52s}
        & 1441
        & \textbf{1}
        & \textbf{1}
        & \textbf{2}
        & Not found
        & 657 \\\hline

    PHP / unserialize
        & \textbf{0 / 10}
        & \textbf{23m05s}
        & \textbf{1h04m39s}
        & \textbf{1h35m08s}
        & 6285
        & \textbf{1}
        & \textbf{1}
        & \textbf{1}
        & Not found
        & 974\\\hline

    Poppler
        & \textbf{0 / 10}
        & \textbf{11m28s}
        & \textbf{49m09s}
        & \textbf{1h33m02s}
        & 9544
        & \textbf{5}
        & \textbf{6}
        & \textbf{8}
        & Not found
        & 543 \\\hline

    SQLite3
        & \textbf{0 / 10}
        & \textbf{33m17s}
        & \textbf{1h02m52s}
        & \textbf{2h42m42s}
        & 4705
        & \textbf{20}
        & \textbf{26}
        & \textbf{31}
        & Not found
        & 226 \\\hline

\end{tabular}
   \begin{flushleft}
      * Two variants of initial fuzzing seeds were used for Belkin: unspecialized (\textit{U}) and specialized
      (\textit{S}) ones. Variant \textit{U} are the default AFL++ seeds for HTTP servers, with which the
      backdoor could never be triggered by AFL++ in 10 runs of 8 hours. Variant \textit{S} are
      specialized seeds, targeting the URL parser of the server, with which the backdoor was
      triggered in 7 of the 10 AFL++ runs. The oracle could always recognize the
      backdoor, once AFL++ had triggered it.
  \end{flushleft}\vspace{-6mm}
\end{table*}

\subsubsection{Duration of phase 1} 
\begin{figure}
    \centering
    
\begin{tikzpicture}[font=\footnotesize]
    \pgfplotsset{
compat = 1.3,
        height=5cm,
        width=\columnwidth - \columnsep,
    }

    \begin{axis}[
        axis y line* = left,
        xtick = {60, 300, 600, 900, 1200},
        xticklabels = {1m, 5m, 10m, 15m, 20m},
        x tick label style = { rotate = 45 },
        extra x ticks = {30},
        extra x tick labels = {30s},
        extra x tick style = { xticklabel style = { yshift = 5pt }},
        xlabel = Phase 1 duration,
        xmin = 0,
        xmax = 1240,
        ymin = 0,
        ymax = 10,
        ytick distance = 2,
        ylabel = {\textcolor{red}{Inputs}},
    ]
        \addplot[mark = o, red]
        coordinates{
            (30, 7)
            (60, 7)
            (300, 5)
            (600, 4)
            (900, 4)
            (1200, 4)
        }; \label{inputs}
    \end{axis}

    \begin{axis}[
legend cell align = left,
        axis x line = none,
        axis y line* = right,
        xmin = 0,
        xmax = 1240,
        ymin = 0,
        ymax = 180,
        ytick distance = 20,
        ylabel = {\textcolor{blue}{Runs}},
    ]
        \addlegendimage{/pgfplots/refstyle=inputs}\addlegendentry{Manually inspected inputs (avg. per run)}
        \addplot[mark = *, blue]
        coordinates{
            (30, 24)
            (60, 24)
            (300, 28)
            (600, 39)
            (900, 45)
            (1200, 51)
        }; \addlegendentry{Failed runs}
    \end{axis}
\end{tikzpicture}
     \vspace{-6mm}
    \caption{\rosa parameter sweep study for the duration of phase 1. For each duration, we performed a total of 180 runs of 8 hours (10 runs per backdoor in \rosarum, including with specialized seeds for the Belkin backdoor).
}
    \label{fig:parameter-sweep}
    \vspace{-4mm}
\end{figure}
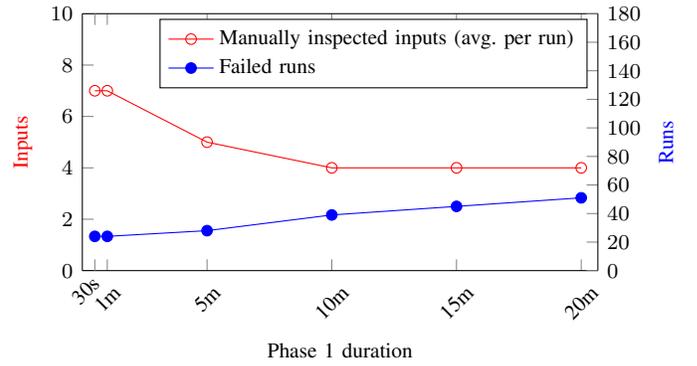

As discussed in \Cref{subsec:rosa-phase-i}, \rosa is parameterized by the duration of the fuzzing campaign at phase 1 and this parameter can impact both the robustness and manual effort of the approach. To evaluate the sensitivity of \rosa to this parameter and pick an optimal value, we have performed a parameter sweep study for durations between 30 seconds and 20 minutes. Within this range, we have measured the average number of manually inspected inputs and the total number of failed runs, for all backdoors in \rosarum. The results are detailed in \Cref{fig:parameter-sweep}. As expected, the number of inspected inputs decreases and the number of failed runs rises when longer durations are used, as this reduces family subsampling but also increases backdoor contamination. In the worst cases, \rosa still remains useful, as only 7 inputs must be analyzed manually, while 72\% of the runs still succeed. As we value better detection capabilities over lower manual effort, we choose \BestPhaseOneDuration as the optimal value and it is the one used for collecting the detailed \rosa evaluation results discussed in the next paragraphs.
\label{subsubsec:parameter-sweep}

\subsubsection{Results discussion} The detailed \rosa evaluation results are presented in \Cref{tab:results-8h}. 
\label{subsubsec:results-discussion}

In terms of \textbf{robustness}, \rosa was able to discover the backdoor during
\SuccessfulRuns (\SuccessfulRunsPercentage) of all \TotalRuns performed runs. For
\BackdoorsWithAllSuccessfulRuns (\BackdoorsWithAllSuccessfulRunsPercentage) of the \BackdoorsTotal
backdoors, all runs were successful. For \BackdoorsWithHalfSuccessfulRuns
(\BackdoorsWithHalfSuccessfulRunsPercentage) of them, at least half of the runs were successful. A
single backdoor was never detected by \rosa using the default seed corpus: the Belkin backdoor. Yet, with specialized seeds (bare GET requests, specifying only a target URL), aiming at the URL parser of the analyzed HTTP server (as URLs are a convenient place to hide backdoor keys), \rosa detected the
backdoor in 7 out of 10 runs, showing that targeting exposed parts of the input space can help
accelerate backdoor detection. 

Experimental results show that the \rosa oracle is robust, with a very low rate of false negatives.
A backdoor miss can be caused either by a fuzzer miss (AFL++ does not generate any input triggering
the backdoor during the run) or a \rosa oracle miss (AFL++ does generate such inputs, but they are
not recognized by the \rosa oracle, i.e., false negatives). By relying on the \rosarum ground
truth, we were able to establish that all of the \FailedRuns failed runs (\num{100}\%) were caused by
a fuzzer miss and none by a \rosa oracle miss. 
In particular, the backdoor contamination effect discussed in \Cref{subsec:rosa-phase-i}
had no influence on detection capabilities in practice.
Backdoor contamination at phase 1 occurred only during \BackdoorContaminationRunsPercentage of all \TotalRuns runs. Moreover, in none of these runs were the backdoor-triggering inputs from phase 2 always matched up with contaminated inputs triggering identical system calls, meaning that backdoor detection was never prevented by contamination.

The main limiting factor for \rosa robustness is the best-effort nature of fuzzing. Like with all
dynamic analyses, detection is limited to the inputs that the fuzzer can test with available
resources. All of the well-known obstacles to fuzzing will increase the probability of a backdoor
miss. These include large input spaces, complex input formats, slow execution times, and
hard-to-inverse operations, like hashing or cryptography (as in the xz~\cite{CVE-2024-3094}
backdoor). How combining \rosa with static analyses could help alleviate the impact of these obstacles (e.g., through slicing~\cite{weiser-1984}), or identify suspicious obstacles, are interesting directions for future work.

In terms of \textbf{speed}, the average time to the first backdoor detection across all 
\SuccessfulRuns successful runs is \RosaAverageDetectionTime.
\BackdoorsWithAverageDetectionTimeBelowOneHour backdoors
(\BackdoorsWithAverageDetectionTimeBelowOneHourPercentage) can be detected by \rosa in less than
one hour on average; \BackdoorsWithAverageDetectionTimeBelowFiveHours backdoors
(\BackdoorsWithAverageDetectionTimeBelowFiveHoursPercentage) in less than five. Here again, the
detection speed of \rosa is principally determined by the velocity at which the fuzzer can trigger
the backdoor. The observed detection times are in line with those of classical fuzzing, where
AFL++ aims for crash-triggering bugs. 

In terms of \textbf{automation level}, for \BackdoorsWithAverageOneInput
(\BackdoorsWithAverageOneInputPercentage) of the \BackdoorsTotal backdoors,
at most one input must be manually inspected on average; for \BackdoorsWithAverageTenInputs
(\BackdoorsWithAverageTenInputsPercentage) of them, at most \num{10} inputs; for the
\BackdoorsWithAverageMoreThanTenInputs (\BackdoorsWithAverageMoreThanTenInputsPercentage) remaining
backdoors, a maximum of
\num{31} inputs had to be manually inspected. These results demonstrate the selectivity of our metamorphic oracle, which cuts by two to four orders of magnitude the number of vetted inputs, compared to the baseline detailed in \Cref{tab:results-8h}, where all AFL++ seeds would have to be vetted. Our experiments also reveal that the semi-automated backdoor vetting procedure
from \Cref{subsec:rosa-post} is effective and requires about 2 minutes of manual effort per input on
average.
We conclude that, for most backdoors, \rosa has an either negligible or moderate manual overhead
(from 2 minutes to 15 minutes) compared to classical fuzzing.

\begin{tcolorbox}[breakable,pad at break*=2mm,
  colback=gray!10,colframe=black,notitle,boxrule=0.5pt,left=2pt,
right=2pt,
top=2pt,
bottom=2pt]
\textbf{\underline{Answer to RQ1}} (usability and usefulness)

\smallskip
\rosa \textbf{detects all backdoors} from our benchmark with a \textbf{level of
    robustness and speed similar to traditional fuzzing}. \rosa also has a \textbf{level of automation similar to traditional fuzzers}, but produces false positives that have to be manually
discarded. Yet, the required manual effort is low, and limited to vetting an average of \RosaAverageInputs suspicious runtime behaviors on our benchmark.  

\end{tcolorbox}

\subsection{RQ2: comparison with the state of the art}
\label{subsec:eval-sota}

\subsubsection{Availability of competing tools} To the best of our knowledge, the state of the art in program analysis for backdoor detection consists of four tools (2013--2017): \textsc{Weasel}~\cite{schuster-2013}, \textsc{Firmalice}~\cite{shoshitaishvili-2015}, \textsc{Stringer}~\cite{thomas-stringer-2017}, \textsc{HumIDIFy}~\cite{thomas-humidify-2017}. After verification and communication with authors, it appears that only \textsc{Stringer} is still both available and working on modern systems.
We compare with \textsc{Stringer} below and discuss the other three as a part of related work (\Cref{sec:related-work}).

\subsubsection{Comparison with \textsc{Stringer}} 
\textsc{Stringer} statically analyzes a binary program and extracts a list of statically-coded strings, likely to be part of a backdoor trigger.
We compare \textsc{Stringer} with \rosa on the \rosarum benchmark.
Among the three case studies from the \textsc{Stringer} paper~\cite{thomas-stringer-2017} that lead to the detection of hard-coded credentials, we could not include in \rosarum the RaySharp and QSee backdoors, as they affect firmware from very old CCTV and DVR devices, which we could not make run on modern systems.
While we were able to run the code involved in the third case study (TRENDnet firmware), we did not include it in \rosarum either, as manual analysis revealed that it was actually not a backdoor, but just a hack to deal with user-defined credentials configured to be empty (the \textsc{Stringer} paper refers to it not as a backdoor but as ``additional functionality'').
As \textsc{Stringer} was also used to recover the FTP command set from a safe version of vsFTPd, we included a different version of this program, infected by an authentic backdoor from another source~\cite{evans-2011}, into \rosarum. 

For each backdoor, we report in \Cref{tab:results-8h} (1) whether a part of the backdoor trigger was detected, (2) how long the detection took and (3) how many strings had to be manually inspected before making a decision about backdoor presence. 

In terms of \textbf{robustness}, only \StringerDetectedBackdoors backdoors out of \BackdoorsTotal
(\StringerDetectedBackdoorsPercentage) can be detected by
\textsc{Stringer}, compared to 17 (\num{100}\%) with \rosa. In particular, only 2 out of the 7 authentic backdoors from \rosarum could be detected by \textsc{Stringer} (vs 7 for \rosa) and the vsFTPd backdoor was detected by \rosa but not by \textsc{Stringer}. These shortcomings are due to \textsc{Stringer} relying on imprecise heuristics. These hypothesize that backdoor triggers should involve static strings with specific properties, which is not true for most non-trivial backdoors. 
In terms of \textbf{speed}, \textsc{Stringer} is \StringerAverageTimesFasterThanRosa times faster than \rosa on average. This is due to \textsc{Stringer} relying on a simple static analysis, where \rosa relies on brute-force dynamic analysis.
In terms of \textbf{automation}, \textsc{Stringer} requires the manual inspection of
\StringerAverageInputs strings on average, compared to \RosaAverageInputs inputs for \rosa. The manual inspection time per unit
appears more important with \textsc{Stringer}, as the expert must reverse-engineer the (binary)
program to locate each string and evaluate its dangerousness.

\begin{tcolorbox}[breakable,pad at break*=2mm,
  colback=gray!10,colframe=black,notitle,boxrule=0.5pt,left=2pt,
right=2pt,
top=2pt,
bottom=2pt]
\textbf{\underline{Answer to RQ2}} (comparison with competing tools)

\smallskip
Out of the four existing program analyzers for backdoor detection, \textbf{only \textsc{Stringer}
    is available and working}. It relies on a simple static analysis that \textbf{cannot detect
    most backdoors from our benchmark}. \textsc{Stringer} identifies the few detected backdoors
    way \textbf{faster than \rosa} but returning \textbf{\StringerAverageTimesMoreInputsThanRosa
    times more, harder-to-vet false positives}.
\end{tcolorbox}

\subsection{Threats to validity}
\label{subsec:eval-threats}

A {first class of threats to the internal validity of our answers to the research questions 
arise because of \emph{possible defects in the software artifacts and manual operations} that we have relied on.
However, AFL++ is a popular, community-maintained and open-source fuzzer, which is employed as a baseline in many fuzzing papers. Our \rosa layer and our experimental infrastructure have been tested at unit level and on small-scale fuzzing campaigns. Our \rosarum benchmark has been tested by one developer and fuzzed by another one. The results of our manual vetting campaigns have been cross-checked with the ground-truth provided by \rosarum.
A second class of threats to internal validity is that our experimental results might not be
significant, due to the \emph{highly random nature of the fuzzing process}. However, we have
mitigated this threat by averaging the results over 10 independent 8-hour runs. In addition, we have systematically discussed the robustness of the \rosa oracle and underlying fuzzer.

Common to all empirical
studies, this one may be of \emph{limited generalizability}. To reduce this
threat, we have performed our experiments over the \BackdoorsTotal backdoors from the \rosarum
benchmark. This includes \BackdoorsAuthentic authentic backdoors, obtained after painstaking
collection and porting efforts, and \BackdoorsSynthetic synthetic backdoors, injected into a
standard fuzzing benchmark, in the style of clean-room design.
In particular, several of these backdoors trigger only system calls as common as benign inputs do, but are still detected by \rosa, showing that the approach is not limited to detecting exotic behaviors.
While one cannot rule out the possibility of an advanced backdoor trigger escaping detection by the \rosa oracle, \rosa would still significantly raise the bar for attackers, likely forcing them to spend extra efforts building more intricate (and thus more noticeable) backdoors.

 \section{Related work}
\label{sec:related-work}

\textbf{Backdoors}. Code-level backdoors induce deviant behaviors that introduce vulnerabilities in software.
Some previous works considered backdoors in other (non-software) parts of computer systems, while others investigated different kinds of malicious or exploitable deviant behaviors in classical software.
The first class of works covers taming backdoors in hardware~\cite{tehranipoor-2010}, cryptography~\cite{easttom-2018}, and machine learning~\cite{li-2024}.
The second includes detecting malware~\cite{aslan-2020} (malicious active programs), as well as exploitable bugs, with a significant part of fuzzing research focusing on detecting memory bugs in memory-unsafe languages~\cite{godefroid-2020}, which pose security threats.

Exploitable bugs can serve a similar purpose as backdoors. So-called ``bugdoors''~\cite{thomas-2018} are bugs injected \emph{on purpose} in programs, for later exploitation by informed attackers. Bugdoors can be more plausibly denied by their authors than backdoors, but they can be detected using well-established tools and practices for software hardening. Bug detection with fuzzing usually relies on simple oracles, like crash detection~\cite{zalewski-2016, fioraldi-2020} and sanitizers~\cite{song-2019}, but metamorphic oracles have also been coupled with fuzzers to find logical bugs in code processors or constraint solvers~\cite{yao-2021, emi-project, rigger-2020, lascu-2021, sqlancer}.

\textbf{Detection of code-level backdoors}. Research on code-level backdoors has been rather scarce~\cite{thomas-2018}. Four approaches and corresponding tools have been proposed to analyze (binary) program code and detect backdoors. We have discussed~\textbf{\textsc{Stringer}} and compared it experimentally to \rosa in \Cref{subsec:eval-sota}. 
\textbf{\textsc{Weasel}}~\cite{schuster-2013} aims at detecting authentication bypass (e.g.,
hardcoded credentials) and hidden commands in protocol binary implementations, like an FTP or SSH
server. \textsc{Weasel} tests the binary with inputs generated from the protocol specification and analyzes the resulting execution traces, to locate the functions or code blocks in the code that process commands or grant authentication. The system and external library calls performed in located code blocks must then be extracted with a disassembler and manually inspected for suspicious patterns. In comparison, \rosa is not restricted to authentication bypass and hidden commands and can detect backdoors in diverse kinds of programs. In addition, \rosa returns dubious inputs and the suspicious system calls that they issue, where \textsc{Weasel} only returns sensitive basic blocks or functions in a binary, to be manually reverse-engineered.
\textbf{\textsc{Firmalice}}~\cite{shoshitaishvili-2015} requires to be supplied with a target point in a binary program, corresponding to the execution of operations restricted to authenticated users. Symbolic execution is then applied on a backwards slice of the program to automatically discover inputs reaching the target, hinting at the possible presence of an authentication bypass in the code. Contrary to \rosa, \textsc{Firmalice} is thus limited to authentication bypass detection and requires manually reverse-engineering a binary to identify the targets.
\textbf{\textsc{HumIDIFy}}~\cite{thomas-humidify-2017} provides a machine learning model, trained
to infer which common protocol (like HTTP or SSH) is implemented by a binary program. The tool
comes with a platform, enabling human experts to specify and verify the feature profile of such
protocols (e.g., ``an HTTP server uses TCP, but not UDP and may read/write files''). Given a binary
program to vet, the machine learning model detects the implemented protocol and the platform
verifies the binary for a divergence with its expected profile. Contrary to \rosa,
\textsc{HumIDIFy} is thus limited to detecting simple hidden features in binaries implementing common protocols. In particular, malicious but profile-compatible behaviors, like hard-coded credentials, cannot be detected by \textsc{HumIDIFy}.
Moreover, vetting if the reported profile violations are actual backdoors is likely to require reverse-engineering the analyzed binaries.
Overall, \textbf{\rosa appears to be the first program analyzer for backdoor detection that:} (1) relies on graybox fuzzing, (2) does not restrict by nature the type of backdoors that can be identified or the category of programs that can be analyzed, and (3) does not require systematic manual reverse-engineering of the analyzed binary.

\textbf{Other works}. The literature on code-level backdoors also includes two other works that are less relevant, but still quite complementary to \rosa.
Ganz et al.~\cite{ganz-2023} detect backdoor injections in collaborative development, with a machine learning model that recognizes suspicious contributions in version control systems repositories.
Thomas et al.~\cite{thomas-2018} present a theoretical framework to define backdoors and reason about detection.
The wider literature includes works that share conceptual elements with \rosa. 
The two-phase approach of \rosa, where one first elucidates standard behaviors and then
searches for behaviors violating these standards, is somewhat similar to the method
of \textsc{diduce}~\cite{hangal-2002}. Yet, \textsc{diduce} elucidates invariants about
the values taken by the expressions of the program, to search for classical
bugs. In contrast, \rosa focuses on input family elucidation to search for backdoors.
Like backdoors, side-channel vulnerabilities provide a subtle way to sneak into
programs, by guessing secret runtime values through observation of non-functional program
behaviors. Complex test oracles were also considered to find
side-channel vulnerabilities with fuzzing, like for example in
\textsc{DifFuzz}~\cite{nilizadeh-2019}. Yet, \textsc{DifFuzz} relies on a differential oracle, where \rosa is based on a metamorphic oracle. In network security, backdoors often refer to maliciously modified server configurations, opening an undocumented remote access to the server. Several works have tried detecting such \emph{network-level backdoors}, like \textsc{pbdt}~\cite{fang-2021}, which uses machine learning to scan Python-based servers.

 \section{Conclusion}
\label{sec:conclusion}

In this work, we have introduced \rosa, the first approach and tool that uses graybox fuzzing to detect code-level backdoors. \rosa complements a state-of-the-art fuzzer (AFL++) with a metamorphic oracle that can identify backdoor triggers at runtime. To enable the experimental evaluation of \rosa and its comparison with existing tools, we have built \rosarum, the first publicly-available benchmark for evaluating backdoor detection tools in diverse programs. \rosarum contains \BackdoorsAuthentic authentic and \BackdoorsSynthetic synthetic backdoors inserted into a standard fuzzing benchmark.
The experimental evaluation of \rosa over the \BackdoorsTotal backdoors of \rosarum shows that
\rosa has a level of robustness, speed, and automation similar to classical graybox fuzzing (for
non-backdoor bugs such as crashes). Compared to the state of the art, \rosa is capable of handling
a wide scope of backdoors and programs and it does not require manually reverse-engineering the
investigated (binary) code.

\ifcameraready 
\section*{Acknowledgment}

We acknowledge the financial support of the French ``Agence Nationale de la Recherche'' (ANR-22-CE39-0012-01) and ``Commissariat à l'Energie Atomique'' (PE-BU-2022-22P1L7). We thank Sébastien Bardin for useful discussions. 
\fi

\clearpage
\IEEEtriggeratref{50}


\end{document}